\documentclass[11pt, twocolumn]{article}

\usepackage{abstract}
\usepackage{lmodern}
\usepackage{graphicx}
\usepackage{epstopdf}
\usepackage[usenames,dvipsnames,svgnames,table]{xcolor}
\usepackage[a4paper]{geometry}
\usepackage[super, comma]{natbib}
\usepackage{fix-cm}
\usepackage{sidecap}
\usepackage{parskip}
\usepackage[squaren,Gray]{SIunits}
\usepackage{amsmath,amsfonts,amssymb}
\usepackage[font={it}]{caption}
\usepackage{subcaption}

\definecolor{navyblue}{RGB}{0,40,150}
\definecolor{linkgreen}{RGB}{0,80,40}
\definecolor{lightgray}{RGB}{150,150,150}
\definecolor{rozovo}{RGB}{150,0, 60}
\definecolor{radinblack}{RGB}{0,0,0}
\usepackage{hyperref}
\hypersetup{
     colorlinks = true,
     citecolor= radinblack,
     linkcolor= radinblack,
     urlcolor= radinblack}

\newcommand*\samethanks[1]{\footnotemark[#1]}
\renewcommand\vec[1]{{\bf #1}}

\newcommand\rev[1]{{#1}}
\newcommand{\dth}{\Delta\theta}

\usepackage{lineno}

\nolinenumbers

\title{\textbf{Contact enhancement of locomotion in \\ \vspace{-0.3cm} spreading cell colonies }\vspace{-0.5cm}}
\date{}
\author{Joseph d'Alessandro%
\thanks{$<$To whom correspondence should be addressed. E-mail: joseph.d'alessandro@ijm.fr,
charlotte.riviere@univ-lyon1.fr$>$}
\thanks{Institut Lumi\`ere Mati\`ere, CNRS UMR 5306, Universit\'e Claude Bernard Lyon 1, Universit\'e de Lyon, Lyon, 69622 Villeurbanne Cedex, France}
\thanks{Current adress: Institut Jacques Monod (IJM), CNRS UMR 7592 et Universit\'e Paris Diderot, 75013 Paris, France},
~Alexandre Solon%
\thanks{Department of Physics, Massachusetts Institute of Technology, Cambridge, MA 02139, USA},
~Yoshinori Hayakawa%
\thanks{Center for Information Technology in Education, Tohoku University, Sendai 980-8578, Japan},
~Christophe Anjard%
\samethanks{2},\\
Fran\c cois Detcheverry%
\samethanks{2},
Jean-Paul Rieu%
\samethanks{2}
~and Charlotte Rivi\`ere%
\samethanks{1}
\samethanks{2}}

\newcommand{\Dru}{{D_{\mathrm{r1}}}}
\newcommand{\Drd}{{D_{\mathrm{r2}}}}
\newcommand{\lamu}{\lambda_1}
\newcommand{\lamd}{\lambda_2}
\newcommand{\pru}{p_1}
\newcommand{\prd}{p_2}
\newcommand{\tauu}{\tau_1}
\newcommand{\taud}{\tau_2}

\newcommand{\be}{\begin{eqnarray}}
\newcommand{\ee}{\end{eqnarray}}

\newcommand{\vc}[1]{\mathbf{#1}}

\newcommand{\vcu}{\vc{u}}

\newcommand{\C}{C}

\begin{document}
\setlength{\parskip}{2ex plus 0.5ex minus 0.2ex}
\setlength{\columnsep}{0.57cm}

\twocolumn[
\begin{@twocolumnfalse}

\maketitle


\textbf {The dispersal of cells from an initially constrained location
  is a crucial aspect of many physiological phenomena ranging from
  morphogenesis to tumour spreading. In such processes, the way
  cell-cell interactions impact the motion of single cells, and in
  turn the collective dynamics, remains unclear. Here, the spreading
  of micro-patterned colonies of non-cohesive cells is fully
  characterized from the complete set of individual trajectories. It
  shows that contact interactions, chemically mediated interactions
  and cell proliferation each dominates the dispersal process on
  different time scales. From data analysis and simulation of an
  active particle model, we demonstrate that contact interactions act
  to speed up the early population spreading by promoting individual
  cells to a state of higher persistence,
  which constitutes an as-yet unreported contact {\em enhancement} of
  locomotion.  Our findings suggest that the current modeling paradigm
  of memoryless interacting active particles may need to be extended
  to account for the possibility of internal states and
  history-dependent behaviour of motile cells.}\\ 
\\

\newpage
\end{@twocolumnfalse}
]
\saythanks

{\huge U}nderstanding how cell assemblies regulate their motility is a
major challenge of current biophysics.\cite{Travis2011} Indeed,
collective effects in the motion of cells play a crucial role
\textit{in vivo} in processes such as wound healing\cite{Friedl2009},
tumour progression\cite{Friedl2003} or
morphogenesis\cite{Carmona2008}.  In understanding the often intricate
relationship between the behaviours at the cellular level and the
population scale, two basic questions arise: how do cell-cell
interactions alter the properties of individual cell motion? How do
they impact the population dynamics?

The trajectory of a cell crawling on a surface is akin to a correlated
random walk characterized by a persistence time beyond which the
motion becomes diffusive\cite{Selmeczi2008,Li2008}. In the absence of
interactions, this would lead on long time to simple diffusion
dynamics at the colony level, as captured in descriptions based on the Fisher-Kolmogorov-Petrovski-Piskunov (FKPP) equation\cite{FKPP,Simpson2013}. However, the assumption of
non-interacting cells is often
unwarranted\cite{Sengers2007,Marel2014}, as several types of cell-cell
interactions affecting the collective dynamics have been uncovered
experimentally. A first class involves long-range interactions, which
may be mediated by a chemical\cite{Gole2011,Phillips2012} as in quorum
sensing, or by the substrate\cite{Angelini2010}.  A second class
includes short-range contact interactions: volume exclusion, cell-cell
adhesion or contact inhibition of locomotion
(CIL)\cite{Abercrombie1953}, which acts to change the direction of
motion of a cell upon contact with another cell.

Despite their local nature, contact interactions have proven essential
to the collective behaviour of cells, at least for dense assemblies
with density near close-packing.  On the edge of a dense colony,
CIL\cite{Mayor2010} or excluded volume\cite{Dyson2014} combined with a
density gradient acts to bias the motion towards free
space\cite{Marel2014,Serra2012,Nnetu2012}, hence facilitating the
spreading of the colony\cite{Sengers2007,Dyson2014,Yates2015}. This
effect is further reinforced by the tension created by leader cells
through adherens junctions\cite{Serra2012,Sepulveda2013}. In the bulk
of a tissue, force transmission through adherens
junctions\cite{Petitjean2010,Tambe2011} (but also nematic
alignment\cite{Coburn2013,Duclos2014} or simple volume
exclusion\cite{Londono2014}) can lead to coordinated motion over
several cell sizes and induce active jamming and glassy
behaviour\cite{Angelini2011,Park2015,Garcia2015}. The slowing down of
tissue dynamics is especially clear in cell systems dominated by
CIL\cite{Mayor2010}, which reduces the cell
persistence\cite{Vedel2013} or effective speed\cite{Fily2012}.

Here, we investigate collective cell migration at moderate density in
assemblies lacking cell-cell adhesion. In contrast to the high-density
regime, this region has received comparatively much less attention so
far. By studying the spreading dynamics of micropatterned
\textit{Dictyostelium discoideum} cell colonies, we find that
cell-cell contacts enhance the cell persistence, an effect that we
refer to as contact enhancement of locomotion (CEL). This phenomenon
results in a speed-up of the colony spreading upon increasing the
packing fraction and defines a novel kind of interaction, which,
instead of acting instantaneously as a physical force, modifies the
internal state of the moving agents and their subsequent behaviour.

\section*{A highly controlled model of cell colony}

\begin{figure*}[ht!]

\centering
\includegraphics[scale=.6]{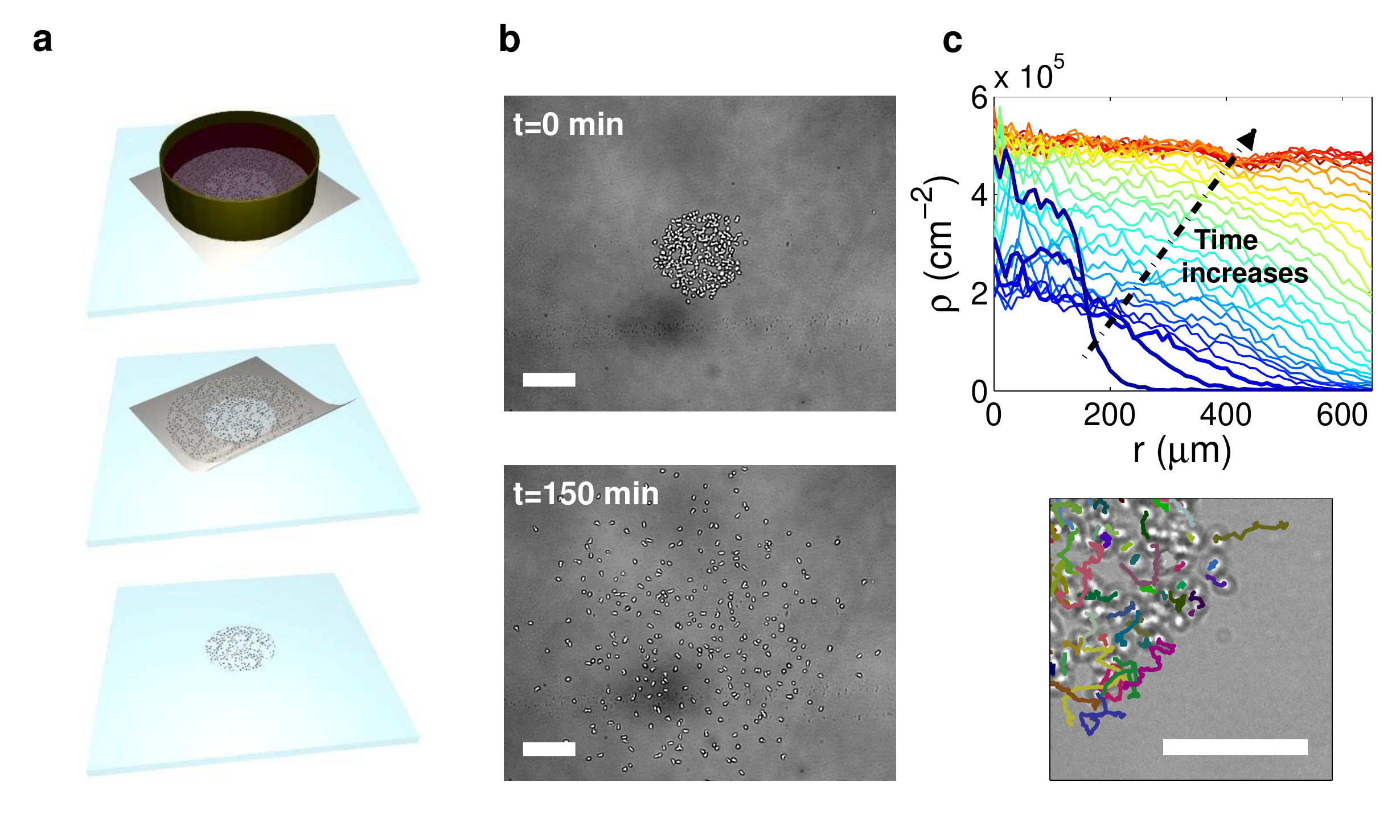}

\caption{A highly-controlled experimental set-up gives full access to colony spreading dynamics at both individual and population scales.\\
  (a) Cartoon of the patterning technique. The cells are first
  deposited in a home-made well (brown) on top of a PDMS stencil
  (light gray) made by soft-lithography techniques (top). After
  $45\,\minute$ of adhesion, the well and the stencil are removed
  (centre), creating an initial circular pattern of
  $\unit{320}\micro\meter$ in diameter (bottom), whose spreading is
  then followed by time-lapse microscopy.  (b) Snapshots of a colony
  with $N_0=245$ cells initially at $t=0\,\minute$ (top) and
  $t=150\,\minute$ (bottom). Scale bars: $200\,\micro\meter$.  (c)
  Top: Evolution of the density profiles $\rho(r, t)$ over $60\,\hour$
  (from blue to red) for one colony with initially $N_0=349$
  cells. All the curves are separated by a $2\,\hour$ interval. The
  first three curves are drawn thicker to highlight the fast initial
  spreading of the colony.  Bottom: Cell trajectories at the edge of
  the initial spot, from $t=0\,\minute$ to $t=60\,\minute$. Scale bar
  $100\,\micro\meter$.}
\label{fig:set-up}
\end{figure*}

We used vegetative
{\it Dictyostelium discoideum} (\textit{D.d.}) cells, which are often considered
as a benchmark for the amoeboid motility of fast-moving cells~\cite{Friedl2001,Friedl2010,Levine2014}.
Moreover, they are especially adapted to study the role of interactions
in the absence of strong cell-cell adhesion, as they do not
form such adhesions in nutrient-rich conditions~\cite{Coates2001}.
To experimentally mimic the dispersal of cells from an initial
location in a reproducible way, we constrained a controlled
number of cells in a disk of diameter $\unit{320}\micro\meter$,
using PDMS micro-stencils
\cite{Poujade2007,Serra2012} (Fig. \ref{fig:set-up}a).  Taking off the
micro-stencil, we let them migrate freely outwards and image the
colony for durations ranging from $\unit{8}\hour$ to $\unit{48}\hour$
(see snapshots in Fig. \ref{fig:set-up}b and Supplementary Movie 1).

We caracterize the colony spreading both at the population and
individual cell levels (Fig. \ref{fig:set-up}c). Making use of the
circular symmetry, field quantities such as the density
(Fig. \ref{fig:set-up}c) are averaged over concentric rings and depend
only on time $t$ and the distance $r$ from the centre of the colony.
At short times, we first observe a decrease of the density in the
centre of the colony as it spreads to invade free space. Then, on time
scales of the order of the doubling time
$\beta^{-1}\sim\unit{9}\hour$, the density starts increasing uniformly
because of cell divisions. Finally, after about $40\hour$ the density
saturates at a carrying capacity
$\rho_{max}\approx 5-8\times
10^{5}\,\text{cell}/\centi\meter^{2}$~(Fig. \ref{fig:set-up}c).

The spreading of the colony is found to be faster in the first few
hours of the experiment. Using the single cell trajectories obtained
by automated cell tracking, this observation can be related to the
average cell speed (see Supplementary Figure 1): After an initial
increase, the cell speed decreases until it reaches a low-motility
plateau at $t\approx\unit{10}\hour$. This behaviour at long times is well explained
by the overall regulation of the motility through a secreted
quorum-sensing factor, which has been evidenced in our group
before~\cite{Gole2011}. Indeed, when repeating the spreading
experiments with a continuous perfusion of fresh medium to rinse out
secreted molecules, the decrease in motility is suppressed and the
colony instead rapidly reaches a high-motility plateau (see
Supplementary Figure 1a). As soon as the perfusion stops, the
concentration of quorum-sensing factors builds up and the cell speed
falls down.

\begin{figure*}[ht!]
\centering
\includegraphics[scale=0.6]{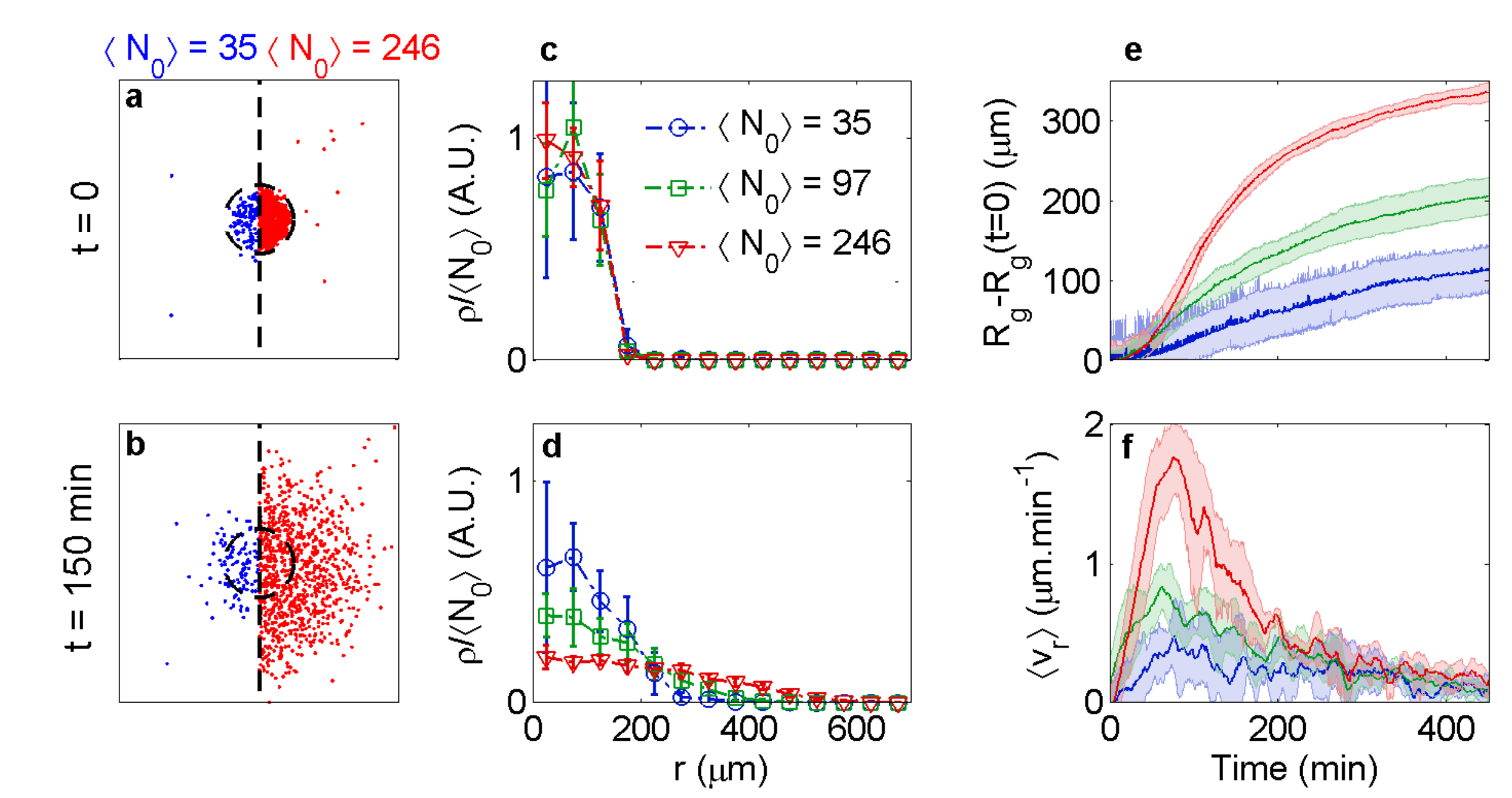}

\caption{Density-dependent colony spreading.\\
  The experiments are divided in three groups to study the effect of
  the initial cell number $N_0$.  (a-b) ``Snapshots'' of the colonies
  for two different groups $N_0=35\pm12$ (blue, left, 208 cells in total) and
  $246\pm66$ (red, right, 1229 cells in total), at $t=0$ and $t=150\,$min. The positions of all cells in each group of
  experiments are represented as coloured points. The dashed circle
  denotes the edge of the stencil.  (c-d) Normalised density profiles
  for each group at $t=\unit{0}\minute$ (c) and $t=\unit{150}\minute$
  (d).  (e) Gyration radius $R_g=\sqrt{\langle r^2\rangle}$ of the
  colonies as a function of time.  (f) Radial velocity
  $\langle v_r\rangle$ as a function of time averaged over the colony
  (same colour code for every panel). The error bars are the standard
  deviation ($n=6, 8, 5$ experiments respectively for
  $N_0 = 35\pm12, N_0=97\pm25, N_0=246\pm66$).}
\label{fig:spread_N0}
\end{figure*}

\section*{Collective effects on the short-time spreading}
We now focus on the short-time spreading of the colony. It strikingly
reveals that the higher the cell density, the faster the colony
spreads (Fig.~\ref{fig:spread_N0}a-b and Supplementary Movies
2-3). This collective effect is seen on the density profiles or on the
gyration radius, $R_g=\sqrt{\langle r^2 \rangle}$, which quantifies
the size of the colony (see Fig.~\ref{fig:spread_N0}c--e). To better
characterise this effect, we compute the radial component $v_r$ of the
velocity of each cell, and average over the colony. We find that this
averaged $\langle v_r\rangle$ exhibits a positive peak around
$\unit{100}\minute$ in the colonies with higher initial density
$N_0 = 246\pm66$ (Fig. \ref{fig:spread_N0}f and
Supplementary Figure 1). This implies that, in this time frame, cells
move outward in average.

The existence of a non-vanishing radial velocity is not surprising in
itself since it is the analogue, for self-propelled
particles~\cite{Cates2013a}, of an outward diffusive flux. However,
one expects the peak to be located at a time of the order of the
persistence time of the particles (see Fig.~\ref{fig:simus}), which is
found around $\tau_p \sim \unit{5}\minute$ for {\em D.d.} cells in
similar conditions~\cite{Gole2011,Bosgraaf2009}. On the contrary, the
radial velocity peak happens here on a much longer time scale
$\sim \unit{100}\minute$. It strongly suggests that an unknown effect
speeds up the spreading on this longer time scale. Importantly, this
effect neither originates from cell division (it happens on a
time-scale much shorter than the doubling time) nor from distant
chemically-mediated interactions arising from bulk soluble molecules
(see Supplementary Figure~\ref{fig:sup_chem}) or from deposited trails
(see Supplementary Figure~\ref{fig:sup_trails}).

Most interestingly, we find that the amplitude of the peak in radial
velocity strongly depends on the number of cells in the colony
(Fig.~\ref{fig:spread_N0}f).  Thus, the spreading rate is collectively
increased through local interactions likely occurring when cells are
in contact. To understand this density-dependent spreading dynamics,
we now turn to a more refined analysis of the motion of individual cells.

\section*{Cell-cell contacts increase their persistence}
\begin{figure*}[ht!]
\centering
\includegraphics[scale=0.7]{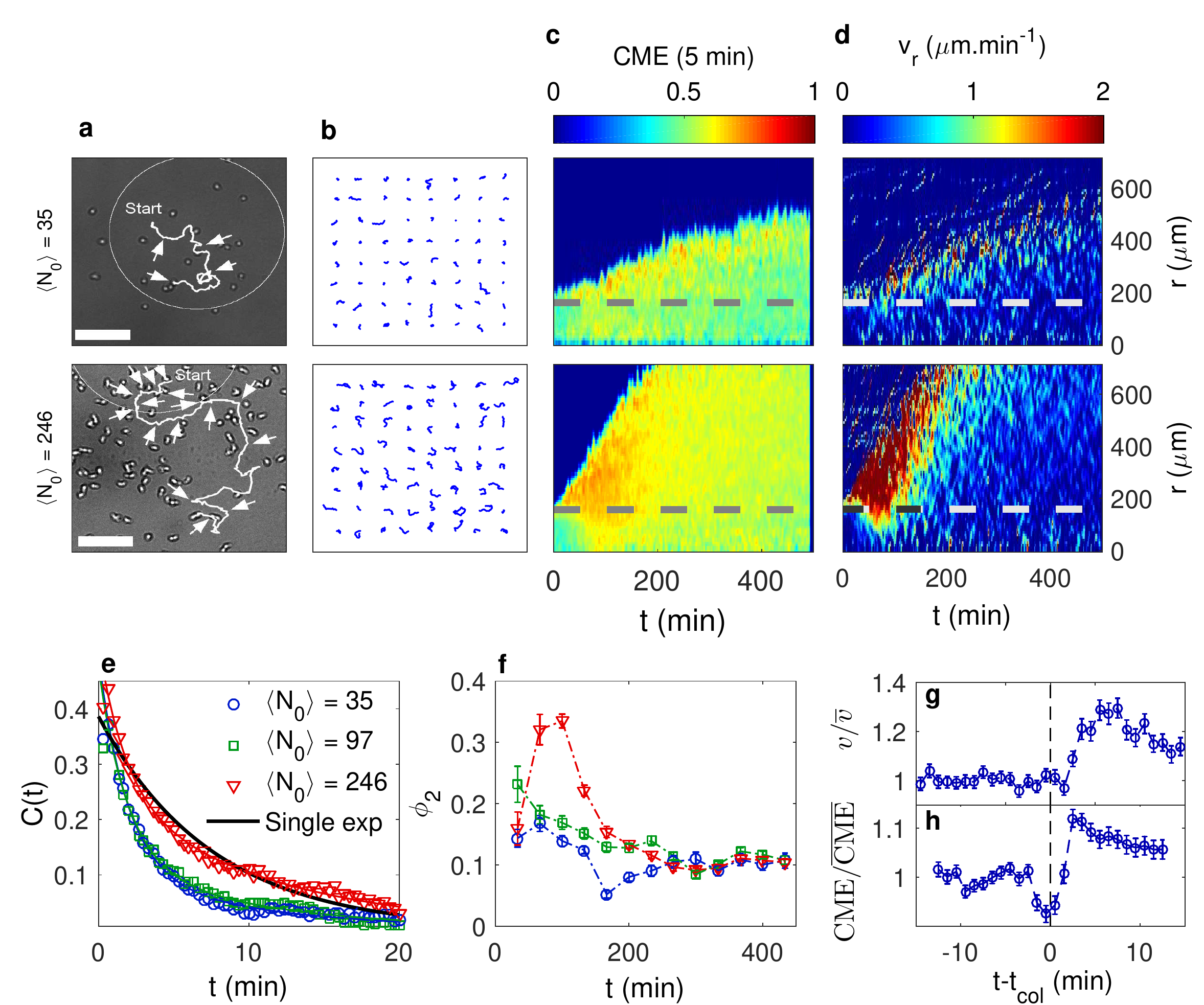}

\caption{Local interactions between cells lead to an increase in cell persistence.\\
  (a--d) Comparison of trajectory properties at low
  ($\langle N_0\rangle=35$, top row) and high
  ($\langle N_0\rangle=246$, bottom row) initial densities. (a)
  Typical cell trajectory (plotted over $\unit{100}\minute$), with
  ``Start'' denoting the initial cell position, the white circle
  representing the border of the initial pattern and the arrows
  pointing to collision events; (b) 64 examples of trajectories from
  $t=100\,\minute$ to $t=130\,\minute$, picked randomly over all cell
  trajectories; (c) spatio temporal dynamics of the CME with
  $\Delta t=\unit{5}\minute$ and (d) of the radial velocity $v_r$ (the
  dashed lines represent the border of the initial colonies).  (e)
  Velocity direction autocorrelation function at $t=67-133\,\minute$
  for various $\langle N_0\rangle$ (symbols, same colour code for
  panels e and f) and fits using the
  expression detailed in Supplementary Information (solid lines) with
  $D_{r1}^{-1}=2\,\minute$ and $\tau_2=10\,\minute$. The thick black line represents the best single-exponential (`Single exp') fit for $\langle N_0\rangle=246$, which misses the experimental
  behaviour at both very short and long times.
(f) Proportion $\phi_2$ of
  cells in mode $2$ as extracted from the fit of the correlation
  functions (the lines are guides for the eye). The error bars represent the 95\% confidence
  interval for the fit parameter.  \rev{(g--h) Normalised speed
    (g) and CME computed with $\Delta t=5\,$min (h) for cells undergoing a
    single collision at $t=t_{col}$ within a 30\,min
    interval. $\overline{v}$ (resp. $\overline{\mbox{CME}}$), denote
    the basal speed (resp. CME) before collision.
The error bars show the SEM for the $n=464$ pieces of trajectory from 232 collisions.}}
\label{fig:persistence}
\end{figure*}

From our dataset of trajectories, we find that the short-time speed-up
of colony spreading is controlled by a transient increase in the
persistence of the cells, the effect being more pronounced the
denser the initial colony. This statement, illustrated by sample
trajectories in Figure \ref{fig:persistence}a-b and Supplementary
Figure~\ref{fig:sup_trajs}, is motivated by a body of quantitative measurements.

First, the cells appear more elongated, hence more polarised, in
denser colonies (Fig. \ref{fig:persistence}a). It is quantified by
computing the cell contours' eccentricity, which increases with
density at early times before relaxing to values corresponding to more isotropic cell shapes (see Supplementary Figure~\ref{fig:sup_ecc}).

Second, we used the coefficient of movement efficiency (CME, see
Methods) to estimate the persistence of the trajectories with good
time and space accuracy. This quantity, for a given interrogation time
$\Delta t$, ranges from $0$ for a motion with persistence time much
smaller than $\Delta t$ to $1$ for ballistic motion. From the
spatio-temporal evolution of the CME measured with
$\Delta t=5\,\minute$, it is clear that the persistence increases with
density (Fig.~\ref{fig:persistence}c). This is especially pronounced
at short times and near the periphery of the colony, where the radial
velocity map also exhibits high values (Fig. \ref{fig:persistence}d).

Finally, another quantitative characterisation is provided by the
velocity direction autocorrelation function
$C(t'-t)=\langle \vcu(t')\cdot \vcu(t)\rangle$, where $\vcu$ is the direction of
motion of a cell. The simplest models of persistent motion
(Ornstein-Uhlenbeck, active Brownian particle or run-and-tumble
motion) all lead to an autocorrelation function which decays
exponentially over the persistence time. In contrast, our data is
better described by a sum of two exponentials
(Fig.~\ref{fig:persistence}e): \be \C(t) = c\, e^{-\gamma
    t} + c'\, e^{-\gamma' t}.
\label{eq:corr_biexp}
\ee  This form could equally arise because of two distinct
populations of cells with different persistence time or because of a
bimodal motion of individual
cells~\cite{Potdar2010,Maiuri2015,Metzner2015}.  \rev{However,
  the first possibility can be ruled out since the relative weights
  $c$ and $c'$ are not constant in time. }

  Thus, all experimental clues point to the fact that each cell is
  able to increase its persistence upon collision. For modeling
  simplicity, although we cannot rule out completely a continuous
  change, we consider cells that switch between two modes of motion, a
  mode~1 whith persistence time is $\Dru^{-1}$ and average duration
  $\tauu$, and a mode~2 with $\Drd < \Dru$ and $\taud$. We used this
  model to fit all experimental data, as explained in the
  Supplementary Information, assuming for simplicity that the mode of
  higher persistence is ballistic
  ($\Drd=0$). Fig.~\ref{fig:persistence}e shows representative
  examples for this fitting procedure.  As an output, we obtain the
  estimates $\Dru^{-1}=$2 min and $\taud=$10 min (see Supplementary
  Figure \ref{fig:sup_R2}), and the fraction
  $\phi_2(t)=\tau_2/\left(\tau_1+\tau_2\right)$
  of cells in mode 2
  (Fig.~\ref{fig:persistence}f). Like the radial velocity, $\phi_2(t)$
  reaches a maximum at around $100\,\minute$, and exhibits an overall
  increase with density, showing that higher densities promote
  switching to the persistent mode.

  Because this enhancement of persistence is not chemically mediated
  (see Supplementary Figs.~1 and~2) and depends strongly on the
  density (Figs.~\ref{fig:spread_N0} and~\ref{fig:persistence}), we
  conclude that contacts -- understood here as collisions or
  short-range interactions -- are the primary cause for the
  phenomenon. Paralleling the CIL acronym, hereafter we refer to this
  effect as CEL, or contact {\em enhancement} of locomotion.  \rev{ To
    confirm more directly the existence of this phenomenon, we look at
    the statistics of cell-cell contacts.  As described in the
    Supplementary Information, we retain only ``clean'' contacts
    (between cells undergoing no other collision for $15\,\minute$
    before and after). We compare the speed and CME of each single
    cell before and after collision and then average over all
    available data (see Fig.~\ref{fig:persistence}g--h, and
    Suplementary Figures~\ref{fig:sup_coll_speed}--\ref{fig:sup_coll_CME_simus_xp}). Both computed quantitites exhibit a
    significant transient increase, demonstrating that CEL is indeed
    responsible for the density-dependent spreading rate of the
    colonies. Note that the analysis of the angular scattering (see Supplementary Figures~\ref{fig:sup_collision}-\ref{fig:sup_angles}) show that the cell-cell contacts have
    no aligning effect likely to promote collective motion and to increase
    the spreading rate by itself.}

\section*{Comparison with minimal active particle models}
To support our experimental findings, we now investigate several
minimal active particle models. Discarding other types of contact
interactions, we show that a collision-induced increase in persistence
is necessary and sufficient to account for the salient features of the
short-time dynamics of the colony spreading. 

\begin{figure*}[ht!]

\centering
\includegraphics[scale=0.6]{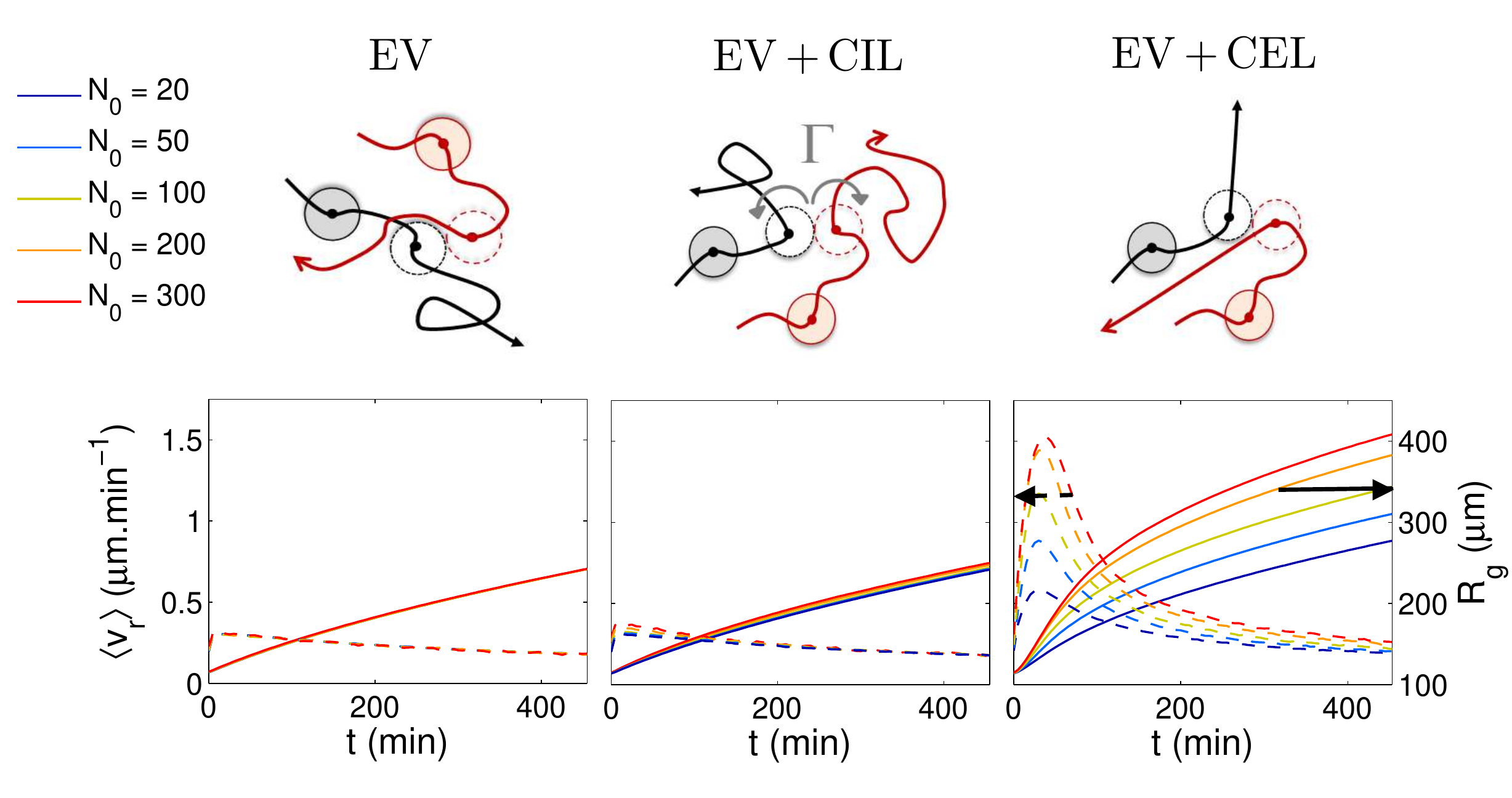}

\caption{Spreading colonies in a particle-based model mimicking the experiments.\\
  Radial velocity $\langle v_r\rangle$ (dashed lines, left axes) and
  colony radius $R_g$ (solid lines, right axes) in simulated colonies
  with different numbers of particles $N_0$ in the three models
  considered. The cartoons on the top row illustrate the different
  contact rules: Excluded-volume only (EV, left), excluded-volume and
  angular repulsion (EV\,+\,CIL, center), excluded-volume and contact
  enhancement of locomotion (EV\,+\,CEL, right). }
\label{fig:simus}
\end{figure*}

Let us consider self-propelled hard disks moving at a constant velocity
$v$. In addition, the direction of motion $\theta$ of a particle is
subject to rotational diffusion with coefficient $D_r$. The motion of
the $i^{\rm th}$ particle is thus governed by the following equations
\begin{linenomath}
\begin{align}
\label{eq:ABPr}\partial_{t}\vec r_i &= v\vec u(\theta_i)+\sum_{j\neq i}\vec f_{ij}(\vec r_i-\vec r_j)\\
\label{eq:ABPtheta}\partial_{t}\theta_i &= \sqrt{2D_r}\eta_i(t),
\end{align}
\end{linenomath}
where $\vec u(\theta_i)=(\cos \theta_i,\sin\theta_i)$ and $\eta_{i}$
is a delta-correlated Gaussian white noise with zero mean and unit
variance. $\vec f_{ij}$ is the steric repelling force exerted by
particle $j$ on particle $i$ (see Methods for more details). As such,
Eqs.~(\ref{eq:ABPr}-\ref{eq:ABPtheta}) describe the active Brownian
particle (ABP) model, well-studied as a minimal model of
phase-separating active particles~\cite{Fily2012,Solon2015}, and a
suitable basis for the modelling of the persistent random motion of
cells. Consistently with the experiments, we take
$v=\unit{5}\micro\meter.\minute^{-1}$ and initially place the particles in
a disk of $\unit{320}\micro\meter$ in diameter. The resulting average
radial velocity and radius of gyration measured in the simulations are
shown as a function of time in Fig.~\ref{fig:simus}.

First, we checked whether this simple interaction rule could yield a
density-dependent spreading. Indeed, if we think of moving cells as
hard-core spheres undergoing a persistent random walk, the
excluded-volume (EV) between the cells gives rise to an outward
pressure~\cite{Solon2015}. This effect is also present for Brownian
hard spheres, for which it can be taken into account by an effective
diffusion coefficient increasing with
concentration~\cite{BrunaChapman}. However, in the present experiments
where cells are relatively sparse -- with packing fractions up to
$0.3$ -- this pressure is not expected to play an important role and
indeed, simulations of Eqs.~(\ref{eq:ABPr})-(\ref{eq:ABPtheta}) with
only hard-core repulsion exhibit no effect of density
(Fig.~\ref{fig:simus}, left).
  
We then further implemented the effect of CIL that is, upon collision,
cells actively reorient away from the contact. To that end, we added
an angular repulsion to the equations of motion, in the form of a
torque \rev{acting on cells undergoing a contact:
  $\Gamma \sum_{j\neq i}H(\|\vec r_{j} - \vec
  r_{i}\|-\sigma_r)\sin(\theta_{i}-\beta_{ij})$
  in equation (\ref{eq:ABPtheta}), where
  $\beta_{ij}=\arg(\vec r_{j} - \vec r_{i})$, $H$ is the Heaviside
  step function implementing the finite radius of interaction
  and $\sigma_r$ is the contact distance (see Methods).} Qualitatively,
one can imagine that by reorienting the direction of motion of
particles toward free space, active reorientation could explain the
experimental data\cite{Mayor2010}. However, even with a large value
$\Gamma=\unit{100}\minute^{-1}$ corresponding to quasi-instantaneous
reorientation, the density-dependent increase in average radial
velocity and spreading rate is an order of magnitude smaller than in
the experiment (Fig.~\ref{fig:simus}, centre). In addition, the (very
limited) peak in radial velocity appears at a very early time, at odds
with the experimental observation. In both situations, the rotational
diffusion was set to $D_r^{-1}=\unit{5}\minute$ to match the average
persistence time of experimental trajectories.

Finally, guided by the experimental observations, we tested a minimal
model of CEL: After each contact, two particles involved in a
collision enter a mode of high persistence (Fig.~\ref{fig:simus},
right). For simplicity, the speed is kept constant and the motion is taken to be ballistic in this
high persistence (mode 2) state.  The particles then relax to the
basal (mode 1) state at a rate $\lambda_2=\tau_2^{-1}$.  We fix
$D_{r1}^{-1}=\unit{2}\minute$ and $\tau_2=10\,\minute$, as determined
from the fit of the experimental data~(see
Fig.~\ref{fig:persistence}). The proportion $\phi_2$ of cells in the
persistent mode (shown in Supplementary Figure~\ref{fig:sup_corr_sim}a) evolves with time
in a density-dependent manner due to changes in the collision
frequency.

We find that this model captures well the collective spreading of the
colony. Indeed, as shown in Fig.~\ref{fig:simus}, the amplitude and
density-dependence of the peak in radial velocity, as well as the
faster increase of $R_g$ at higher density, are well
captured. Overall, given the simplicity of the model, the
agreement between simulations and experiments appears surprisingly
good (see Supplementary Movies 2--5 for a visual comparison).

\section*{Discussion}
We studied the dynamics of spreading colonies of \textit{D.d.}
cells. Using a micro-fabrication technique, we were able to produce
initial colonies with controlled shape and number of cells.  We showed
that the long-time dynamics of the spreading is controlled by cell
divisions and long-range interactions through a quorum-sensing
factor. On the contrary, these two effects are absent from the
short-time dynamics, thus allowing us to study the effect of cell-cell
contacts. We found that cell contacts enhance the spreading of the
colony by increasing the speed and the persistence time of the cells motion. This {\it contact
  enhancement of locomotion} is further supported by a simple active
particle model reproducing the main characteristics of the
experimental data.

The phenomenon of ``CEL'' that we have described here seems \textit{a
  priori} different from CIL, which acts to change the direction of
motion of colliding cells. However, \rev{ they are not mutually
  exclusive (see Supplementary Figure~\ref{fig:sup_coll_CME_simus_xp}) and could even share a common microscopic origin.}%
Indeed, the current explanation for CIL is that the protrusions
driving the motion are inhibited in the contact
region~\cite{Mayor2010,Stramer2016}. Other protrusions can thus
develop elsewhere on the cell's periphery, leading to a new direction
of motion. We could hypothesize that, similarly, \rev{either the
  inhibition of ruffling in the contact region or the stabilisation of
  the new protrusions reinforces the new polarity thereby increasing
  the speed and persistence of the motion. An important difference between the two
  effects is that CEL involves memory while CIL is usually modeled as
  an ``instantaneous'' contact
  process~\cite{Davis2012,Zimmermann2016,Camley2016,Szabo2016}.}
The two could have different relative importances depending on the
situation. For example, in dense cell clusters, the reorientation
induced by CIL forces the particles on the edge to move
outward~\cite{Carmona2008,Zimmermann2016} whereas, as we saw, in
sparser colonies the increase in persistence is the dominating effect.

The importance of contact interactions \textit{in vivo} is not
completely understood. CIL has been found to play an important role in
neural crest migration~\cite{Carmona2008} and the loss of heterotypic
CIL (between different cell types) is thought to be crucial in the
invasion of healthy tissues by cancer
cells~\cite{Friedl2003,Abercrombie1979}. Similarly, CEL could be an
advantage in invading the surrounding environment efficiently, in the
case of a less cohesive group of cells. Such a situation can be
encountered during the escape of highly metastatic and invasive cancer
cells, as well as in cell morphogenesis and microbial dispersal.  More
generally, it should be noticed that CEL is reminiscent of escape
mechanisms, found in various organisms, which involve a temporary
change of the motile behaviour and can lead to surprising collective
effects\cite{Ramdya2015}. 

Finally, the new type of interaction uncovered here opens questions
for active matter. Indeed, it exemplifies the wide range of possible
interactions between active particles, compared to the usual
``physicist's particles'', which can lead to a rich phenomenology. It
opens the door to further studies of interactions that act on an
additional internal degree of freedom, which could exhibit other
interesting effects.

{\textbf{METHODS}} 

{\small
\subsection*{Cell culture}
We used \textit{Dictyostelium discoideum} cells from the strain
AX2. The cells were cultured on cell-culture-treated Petri dishes (BD
Falcon) in HL5 medium with glucose (Formedium) and kept in a
temperature-controlled incubator at 22.5\celsius, with a doubling
time $\beta^{-1}\sim9\,\hour$. Before every experiment,
the cells were detached from the dish, centrifuged $\unit{5}\minute$
at $663g$, harvested and resuspended at the seeding density.

\subsection*{Sample preparation}
A reusable mould on Si wafer comprising an array of squares with
circular pillars of height $\sim\unit{150}\micro\meter$ and diameter
$\unit{320}\micro\meter$ in the centre was fabricated in SU8
photoresist using classical soft lithography techniques and its
surface was silanized to make it non-adherent. PolyDiMethylSulfoxyde
(PDMS, Corning) mixed with curing agent at a 1:10 mass ratio was spin
coated on the mould for $\unit{1}\minute$ at $750$~rpm to a target
thickness of $\unit{70}\micro\meter$. The squares were cut and peeled
off. Usually a thin PDMS membrane obstructed the hole. It was then
removed with a surgical blade under the microscope at low
magnification.

The square stencil was stuck on the ground of a
$\unit{3.5}\centi\meter$ wide culture dish and a homemade small
plastic well was stuck on it using silicon seal. A droplet of medium
was deposited into the well and the sample was placed under vacuum for
$\unit{15}\minute$ to help the medium enter the central hole of the
stencil and wet the dish's surface.

The cell suspension was added in the well and the sample was placed in
the incubator for $\unit{45}\minute$ to let the cells sediment and
adhere. Then, the plastic well and the stencil were removed with
surgical tweezers. Last, the spreading colony was imaged using a
slightly defocused bright-field microscope (TE2000, Nikkon) at 10X
magnification and a wide-field Andor Zyla sCMOS camera. A time-lapse
movie was recorded for up to $\unit{48}\hour$ using MicroManager
software with a $\unit{20}\second$ time-interval, while the
temperature was kept constant at 22.5\celsius.

For perfusion experiments, we designed a macrofluidic chamber by
sealing the culture dish with an adapted cover containing an input and
an output tube. The former was linked to a $\unit{1}\liter$ supply
bottle of fresh HL5 medium under controlled overpressure (OB1
controller, Elveflow) while the latter was linked to a disposal
bottle. All the system was closed sterilely. We used a flow rate of
$\unit{100}\milli\liter\per\hour$ so that the chamber volume of about
$\unit{10}\milli\liter$ was completely renewed every 6 minutes. We
were thus able to maintain a stable medium renewal over 9 hours.

\subsection*{Image processing}
The cells' positions were retrieved using homemade ImageJ macros based
on the 'Find Maxima' built-in function. Then the individual
trajectories were reconstructed with a squared-displacement
minimization algorithm (http://site.physics.georgetown.edu/matlab/)
and the data analysed using homemade Matlab programs.

In particular, the CME was defined as:
\begin{equation}
\text{CME}_{\delta t}(t) = \frac{\|\mathbf{r}(t+\frac{\delta t}{2})-\mathbf{r}(t-\frac{\delta t}{2})\|}{\int_{t'=t-\frac{\delta t}{2}}^{t+\frac{\delta t}{2}}\|\mathbf{v}(t')\|dt'}
\label{CME}
\end{equation}

\subsection*{Simulations}
Simulations were carried out by integrating the Langevin equations
Eqs.~(\ref{eq:ABPr}-\ref{eq:ABPtheta}) using a Euler integration
scheme with time steps $\Delta t=10^{-3}$ min. The hard-core repulsion
between particles is modelled by a Weeks-Chandler-Andersen potential
$V(r)=4\left[ \left(\frac \sigma r \right)^{12}-\left(\frac \sigma r
  \right)^{6}\right]+1$
if $r<2^{1/6}\sigma$ and $0$ otherwise, where $\sigma=10\,\mu m$ is
the particle diameter. We define two particles as being in contact
when their relative distance $r<\sigma_r=2^{1/6}\sigma$.  \rev{ In the
  simulations with CIL, the torque term is turned on only during the
  contacts, when $r<\sigma_r$.} In the simulation including CEL, a
contact triggers a ballistic run which lasts for an exponentially
distributed time with rate $\lambda_2$.}


\bibliographystyle{unsrt}

\begin{thebibliography}{10}
\bibitem{Travis2011}
Travis, J. Mysteries of the cell: Cell biology's open cases.
 \emph{Science} \textbf{334,} 1051 (2011).

\bibitem{Friedl2009}
Friedl, P. \& Gilmour, D. Collective cell migration in morphogenesis, regeneration and cancer.
 {\em Nat Rev Mol Cell Biol} \textbf{10,} 445--457 (2009).

\bibitem{Friedl2003}
Friedl, P. \& Wolf, K. Tumour-cell invasion and migration: diversity and escape mechanisms. \textit{Nature Reviews Cancer} 3\textbf{,} 362-374 (2003).

\bibitem{Carmona2008}
Carmona-Fontaine, C. \textit{et al.} Contact inhibition of locomotion in vivo controls neural crest directional migration. \textit{Nature} \textbf{456,} 7224 (2008).

\bibitem{Selmeczi2008}
Selmeczi, D. \textit{et al.} Cell motility as random motion: A review.
 {\em European Physical Journal: Special Topics} \textbf{157,} 1--15 (2008).
 
\bibitem{Li2008}
Li, L., Norrelkke, S.F. \& Cox, E.C. Persistent cell motion in the absence of external signals: A search  strategy for eukaryotic cells.
 {\em PLoS ONE}, \textbf{3,} 5 (2008).

\bibitem{FKPP}
Kolmogorov, A., Petrovskii, I. \& Piscounov, N. A study of the diffusion equation with increase in the amount of substance, and its application to a biological problem.
  {\em Math. Mech.} \textbf{1,} 1-–25 (1937).

\bibitem{Simpson2013}
Simpson, M.J. \textit{et al.} Quantifying the roles of cell motility and cell proliferation in a circular barrier assay.
 {\em Journal of the Royal Society, Interface / the Royal Society} \textbf{10,} 20130007 (2013).

\bibitem{Sengers2007}
Sengers, B.G., Please C.P. \& Oreffo R.O.C. Experimental characterization and computational modelling of two-dimensional cell spreading for skeletal regeneration.
{\em Journal of the Royal Society, Interface / the Royal Society}, \textbf{4,} 1107--1117 (2007).

\bibitem{Marel2014}
Marel, A.K. \textit{et al.} Flow and Diffusion in Channel-Guided Cell Migration.
 {\em Biophys. J.}, \textbf{107,} 1054--1064 (2014).

\bibitem{Gole2011}
Gol\'{e}, L., Rivi\`{e}re, C., Hayakawa, Y. \& Rieu, J.P. A quorum-sensing factor in vegetative Dictyostelium Discoideum cells revealed by quantitative migration analysis.
 {\em PLoS ONE} \textbf{6,} 1--9 (2011).

\bibitem{Phillips2012}
 Phillips, J. \& Gomer, R. A secreted protein is an endogenous chemorepellant in \textit{Dictyostelium discoideum}.
  {\em Proc. Natl Acad. Sci. USA}, \textbf{109,} 10990--10995 (2012).

\bibitem{Angelini2010}
Angelini, T.E., Hannezo, E., Trepat, X., Fredberg, J.J. \& Weitz, D.A. Cell migration driven by cooperative substrate deformation patterns.
 {\em Phys. Rev. Lett.} {\bf 104,} 168104 (2010).
 
\bibitem{Abercrombie1953}
Abercrombie, M. \& Heaysman, J.E.
Observations on the social behaviour of cells in tissue culture: I. Speed of movement of chick heart fibroblasts in relation to their mutual contacts.
\textit{Experimental cell research} \textbf{5,} 111--131 (1953).

\bibitem{Mayor2010}
  Mayor, R. \& Carmona-Fontaine, C. Keeping in touch with contact inhibition of locomotion.
 {\em Trends in cell biology} \textbf{20,} 319--328 (2010).

\bibitem{Dyson2014}
Dyson, L. \& Baker, R.E. The importance of volume exclusion in modelling cellular migration.
 {\em J. Math. Biol.} {\bf 71,} 679--711 (2014).

\bibitem{Serra2012}
Serra-Picamal, X. \textit{et al.} Mechanical waves during tissue expansion.
 {\em Nat. Phys.} \textbf{8,} 628--634 (2012).
 
\bibitem{Nnetu2012}
Nnetu, K.D., Knorr, M., Strehe, D., Zink, M. \& K\"as, J.A. Directed persistent motion maintains sheet integrity during multi-cellular spreading and migration.
 {\em Soft Matter} \textbf{8,} 6913 (2012).
 
\bibitem{Yates2015}
Yates, C.A., Parker, A. \& Baker, R.E. Incorporating pushing in exclusion-process models of cell migration.
 {\em Phys. Rev. E} \textbf{91,} 052711 (2015).

\bibitem{Sepulveda2013}
Sep\'ulveda, N. \textit{et al.} Collective cell motion in an epithelial sheet can be quantitatively described by a stochastic interacting particle model.
 {\em PLoS Comput. Biol.} \textbf{9,} e1002944 (2013).

\bibitem{Petitjean2010}
Petitjean, L. \textit{et al.} Velocity fields in a collectively migrating epithelium.
 {\em Biophysical J.} \textbf{98,} 1790--1800 (2010).

\bibitem{Tambe2011}
Tambe, D.T. \textit{et al.} Collective cell guidance by cooperative intercellular forces.
 {\em Nat. Mat.} \textbf{10} 469--475 (2011).

\bibitem{Coburn2013}
Coburn, L., Cerone, L., Torney, C., Couzin, I.D. \& Neufeld, Z. Interactions lead to coherent motion and enhanced chemotaxis of migrating Cells.
  {\em Phys. Biol.} \textbf{10,} 046002 (2013).

\bibitem{Duclos2014}
Duclos, G., Garcia, S., Yevick, H.G. \& Silberzan, P. Perfect nematic order in confined monolayers of spindle-shaped cells.
 {\em Soft Matter} \textbf{10,} 2346 (2014).

\bibitem{Londono2014}
Londono C. \textit{et al.} Nonautonomous contact guidance signaling during collective cell migration.
 {\em Proc. Natl Acad. Sci. USA} \textbf{111} 1807--1812 (2014).

\bibitem{Angelini2011}
Angelini, T.E. \textit{et al.} Glass-like dynamics of collective cell migration.
 {\em Proc. Natl Acad.
Sci. USA} \textbf{108,} 4714--4719 (2011).
 
\bibitem{Park2015}
Park, J.-A. \textit{et al.} Unjamming and cell shape in the asthmatic airway epithelium.
 {\em Nat. Mat.} \textbf{14,} 1040--1049 (2015).

\bibitem{Garcia2015}
Garcia, S. \textit{et al.} Physics of active jamming during collective cellular motion in a monolayer.
 {\em Proc. Natl Acad. Sci. USA} \textbf{112,} 15314--15319 (2015).

\bibitem{Vedel2013}
Vedel, S., Tay, S., Johnston, D.M., Bruus, H. \& Quake, S.R. Migration of Cells in a Social Context.
 \emph{Proc. Natl Acad. Sci. USA} \textbf{110,} 129-134 (2013).

\bibitem{Fily2012}
Fily, Y. \& Marchetti, M.C. Athermal Phase Separation of Self-Propelled Particles with No Alignment.
 \emph{Phys. Rev. Lett.} \textbf{108,} 235702 (2012).

\bibitem{Friedl2001}
Friedl, P., Borgmann, S. \& Br\"ocker, E. B. Amoeboid leukocyte crawling through extracellular matrix: lessons from the Dictyostelium paradigm of cell movement.
 {\em Journal of Leukocyte Biology} \textbf{70,} 491-509 (2001). 

\bibitem{Friedl2010}
Friedl, P. \& Wolf, K. Plasticity of cell migration: a multiscale tuning model. {\em J. Cell Biol.} \textbf{188,} 11--9 (2010).

\bibitem{Levine2014}
Levine, H. Learning Physics of Living Systems from {\it Dictyostelium}.
 {\em Phys. Biol.} \textbf{11,} 053011 (2014).

\bibitem{Coates2001}
Coates, J. C. \& Harwood, A. J. Cell-cell adhesion and signal transduction during {\it Dictyostelium development}.
 {\em J. Cell Sci.} \textbf{114,} 4349-4358 (2001).

\bibitem{Poujade2007}
Poujade, M. \textit{et al.} Collective migration of an epithelial monolayer in response to a model wound.
 {\em Proc. Natl Acad. Sci. USA} \textbf{104,} 15988--15993 (2007).

\bibitem{Cates2013a}
Cates, M.E. \& Tailleur, J. When are active Brownian particles and run-and-tumble particles equivalent? Consequences for Motility-induced Phase Separation. \emph{Europhys. Lett.} \textbf{101,} 20010 (2013).

\bibitem{Bosgraaf2009}
Bosgraaf, L. \& Van Haastert, P.J.M. The ordered extension of pseudopodia by amoeboid Cells in the absence of external cues.
 {\em PLoS ONE} \textbf{4,} 4 (2009).

\bibitem{Solon2015}
Solon, A.P. \textit{et al.} Pressure and phase equilibria in interacting active Brownian spheres.
 \emph{Phys. Rev. Lett.} \textbf{114,} 198301 (2015).

\bibitem{BrunaChapman}
Bruna, M. \& Chapman, S.J. Excluded-volume effects in the diffusion of hard spheres.
 {\em Physical Review E} \textbf{85,} 011103 (2012).

\bibitem{Peruani2007}
Peruani, F. \& Morelli, L.G. Self-propelled particles with fluctuating speed and direction of motion in two dimensions.
 {\em Phys. Rev. Lett.} \textbf{99,} 010602 (2007).

\bibitem{Potdar2010}
 Potdar, A.A., Jeon, J., Weaver, A.M., Quaranta, V. \& Cummings, P.T. Human mammary epithelial cells exhibit a bimodal correlated random walk pattern.
 {\em PLoS ONE} \textbf{5,} e9636 (2010).

\bibitem{Metzner2015}
 Metzner, C. \textit{et al.} Superstatistical analysis and modelling of heterogeneous random walks.
 {\em Nat. Commun.} \textbf{6,} 7516 (2015).

\bibitem{Selmeczi2005}
 Selmeczi, D. \textit{et al.} Cell motility as persistent random motion: theories from experiments.
 {\em Biophys. J.} \textbf{89,} 912--931 (2005).

\bibitem{Maiuri2015}
Maiuri, P. \textit{et al.} Actin Flows Mediate a Universal Coupling between Cell Speed and Cell Persistence.
 {\em Cell} \textbf{161,} 374--386 (2015).

\bibitem{Stramer2016}
Stramer, B.A. \& Mayor, R. Mechanisms and in vivo functions of contact inhibition of locomotion.
 {\em Nat. Rev. Mol. Cell. Biol.} \textbf{118} (2016). 

\bibitem{Davis2012}
Davis, J.R., \textit{et al.} Emergence of embryonic pattern through contact inhibition
of locomotion.
 {\em Development} \textbf{139,} 4555-4560 (2012)

\bibitem{Zimmermann2016}
Zimmermann, J., Camley, B.A., Rappel, W.-J. \& Levine, H. Contact inhibition of locomotion determines cell-cell and cell-substrate forces in tissues.
 {\em Proc. Natl Acad. Sci. USA} \textbf{113,} 2660-2665 (2016).

\bibitem{Camley2016}
Camley, B.A., Zimmermann, J., Levine, H. \& Rappell, W.-J. Emergent collective chemotaxis without single-cell gradient sensing.
 {\em Phys. Rev. Lett.} \textbf{116,} 098101 (2016).

\bibitem{Szabo2016}
Szabo, A., \textit{et al.} In vivo confinement promotes collective migration of neural crest cells.
 {\em J. Cell. Biol.} \textbf{213,} 543--555 (2016).

\bibitem{Abercrombie1979}
Abercrombie, M. Contact inhibition and malignancy.
 \textit{Nature} \textbf{281,} 259--262 (1979).

\bibitem{Ramdya2015}
Ramdya, P. \textit{et al.} Mechanosensory interactions drive collective behaviour in \textit{Drosophila}.
 \emph{Nature} \textbf{519,} 233--236 (2015).
 
\bibitem{Heid2004}
Heid, P. J. \textit{et al.} The role of myosin heavy chain phosphorylation in Dictyostelium motility, chemotaxis and F-actin localization.
{\em J. Cell Sci.} \textbf{117,} 4819--4835 (2004).

\end{thebibliography}

\vspace{1cm}


{\textbf{ACKNOWLEDGEMENTS}}

{\small The authors are grateful to R. Fulcrand for his help in micro-fabrication, to V. Hakim for stimulating
  discussions and to C. Cottin-Bizonne for her comments on the manuscript. J.d.A. has been partially supported by the Fondation
  ARC pour la Recherche sur le Cancer and by the Programme d'Avenir
  Lyon-Saint \'Etienne. A.S. acknowledges funding through a PLS
  fellowship of the Gordon and Betty Moor foundation. J.d.A., C.R. and
  J.P.R. belong to the CNRS consortium CellTiss and to the LIA
  ELyTLab.  } \vspace{24pt}

{\textbf{AUTHOR CONTRIBUTIONS}}

{\small J.d.A., J.P.R. and C.R. designed experiments; J.d.A.
  performed experiments and analysed experimental data; J.d.A. and
  A.S. conceived the particle-based models; A.S. performed simulations
  and analysed simulation data; C.A. contributed to design of
  experiments in Supplementary Figure 1 and provided AprA$^{-}$ cells;
  F.D. computed the analytical results on bimodal trajectories and
  helped with the fitting procedure; Y.H. assisted in the data
  analysis and interpretation; J.d.A. and A.S. wrote the manuscript;
  F.D., J.P.R. and C.R. made substantial contribution to the
  manuscript; all authors discussed and interpreted the data, read and
  commented on the manuscript; J.P.R. and C.R. supervised the project.
}

\vspace{24pt}
\newpage

\renewcommand{\figurename}{Supplementary Figure}
\setcounter{figure}{0}
\onecolumn
\begin{center}
{\LARGE\bf Supplementary information\\}
\end{center}

\section{An effect of local interactions}

\begin{figure*}[ht!]

\centering
\includegraphics[scale=0.6]{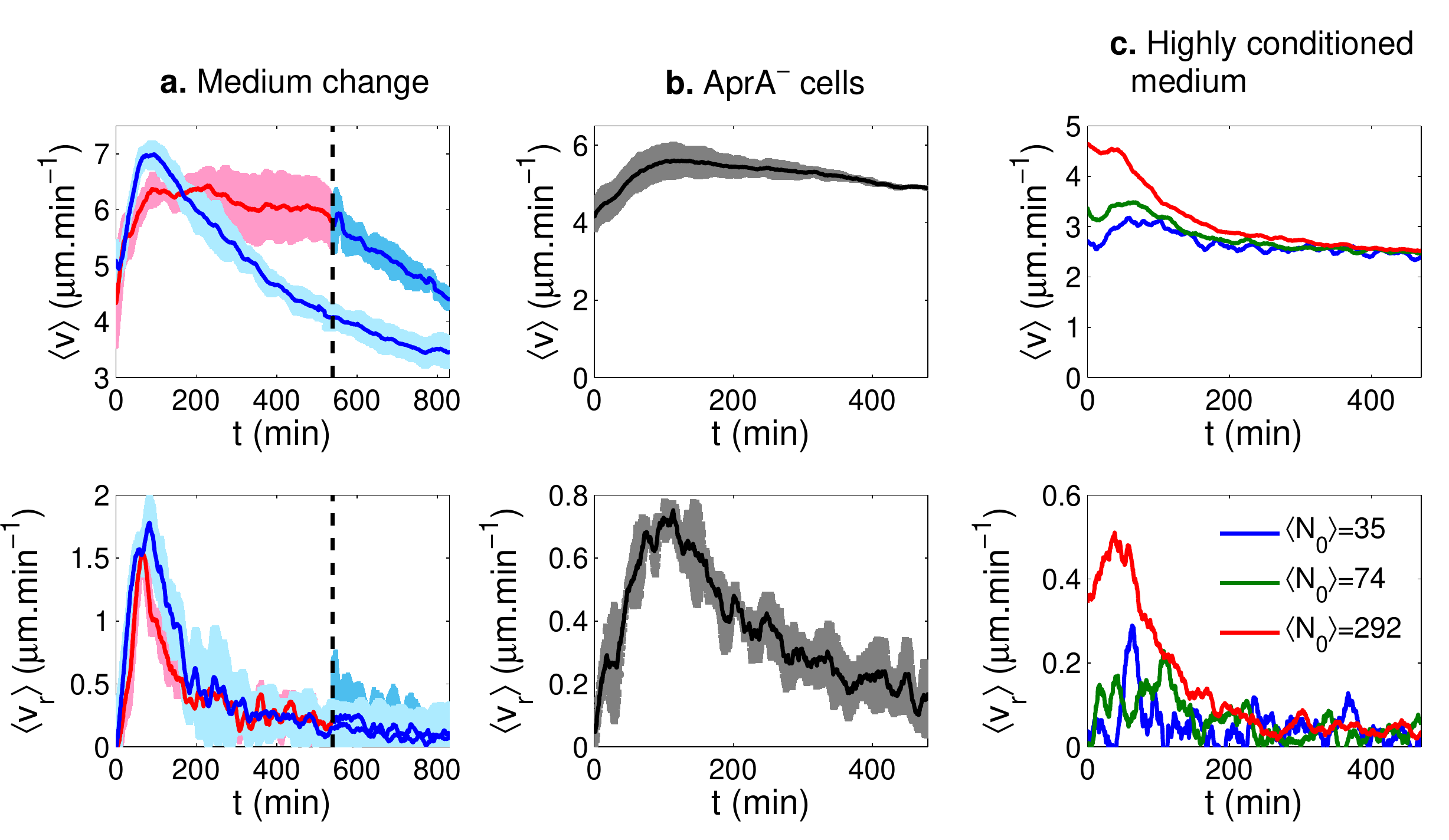}

\caption{The cells' outward motion is not due to a large-scale chemical communication.\\
  Colony-averaged speed (top row) and radial velocity (bottom row) in
  various conditions.  (a) Medium exchange. Control (blue) and samples
  with continuous medium exchange (red). The speed is maintained
  constant until the exchange is stopped (vertical dashed line),
  whereas the radial motion $\langle v_r\rangle$ is robust to medium
  exchange.  (b) $aprA^-$ cells still exhibit outward motion,
  although they do not secrete AprA, the only known endogenous
  chemorepellent in \textit{Dictyostelium discoideum} (c) Cells in
  highly conditioned medium (HCM), prepared by letting cells in
  culture in it before filtering, so that all secreted molecules are
  already at a high concentration. The cell motion is affected but
  there is still a density-dependent peak in $\langle v_r\rangle(t)$.}
\label{fig:sup_chem}
\end{figure*}

\rev{
We checked that long-distance interactions were not responsible for the
enhancement of the colony spreading rate. To that end, we measured the average speed
$\langle v\rangle$ and radial velocity $\langle v_r\rangle$ during the
spreading of cell colonies in three different control conditions, as described
below. We looked especially at the presence of the peak in $\langle v_r\rangle(t)$,
which, in our experiments, is the macroscopic signature of the collective
enhancement of the spreading.} Firstly, we designed a fluidic
system that allowed to continuously change the sample's medium, so
that any secreted (or depleted) molecule was rinsed out (or rescued),
hence preventing any large-scale chemical sensing such as
chemorepulsion or quorum-sensing. It efficiently suppressed the
overall regulation of cell motility through a known quorum-sensing
factor (QSF)~\cite{Gole2011} (Supplementary Figure
\ref{fig:sup_chem}a, top), but did not affect the outward motion
(Supplementary Figure \ref{fig:sup_chem}a, bottom) which shows
  that the collective effect is still present. Secondly, we used
$aprA^-$ cells which do not produce the protein AprA, so far the only
endogenous chemorepellent molecule known for \textit{Dictyostelium
  discoideum}~\cite{Phillips2012}. Although the motility of these
cells is slightly different from the wild-type cells, resulting in
slightly modified colony dynamics, the main effect of outward motion was
still observed (Supplementary Figure \ref{fig:sup_chem}b). Last,
experiments were done in highly conditioned medium (HCM). This medium
is prepared by letting cells in culture in fresh HL5 medium for
typically 2 days, so that it is supplemented with molecules secreted
by the cells, at high concentration. Thus, one would expect the
concentrations in slowly degraded molecules to be above the saturation
of their detection by the cells, hence screening any additional
secretion during the experiment. Although the dynamics is again
modified mainly due to the presence of QSF at high concentration in
HCM~\cite{Gole2011}, the collective effect of outward motion is still
apparent in this situation (Supplementary Figure \ref{fig:sup_chem}c).

\begin{figure*}[ht!]
\centering
\includegraphics[scale=0.85]{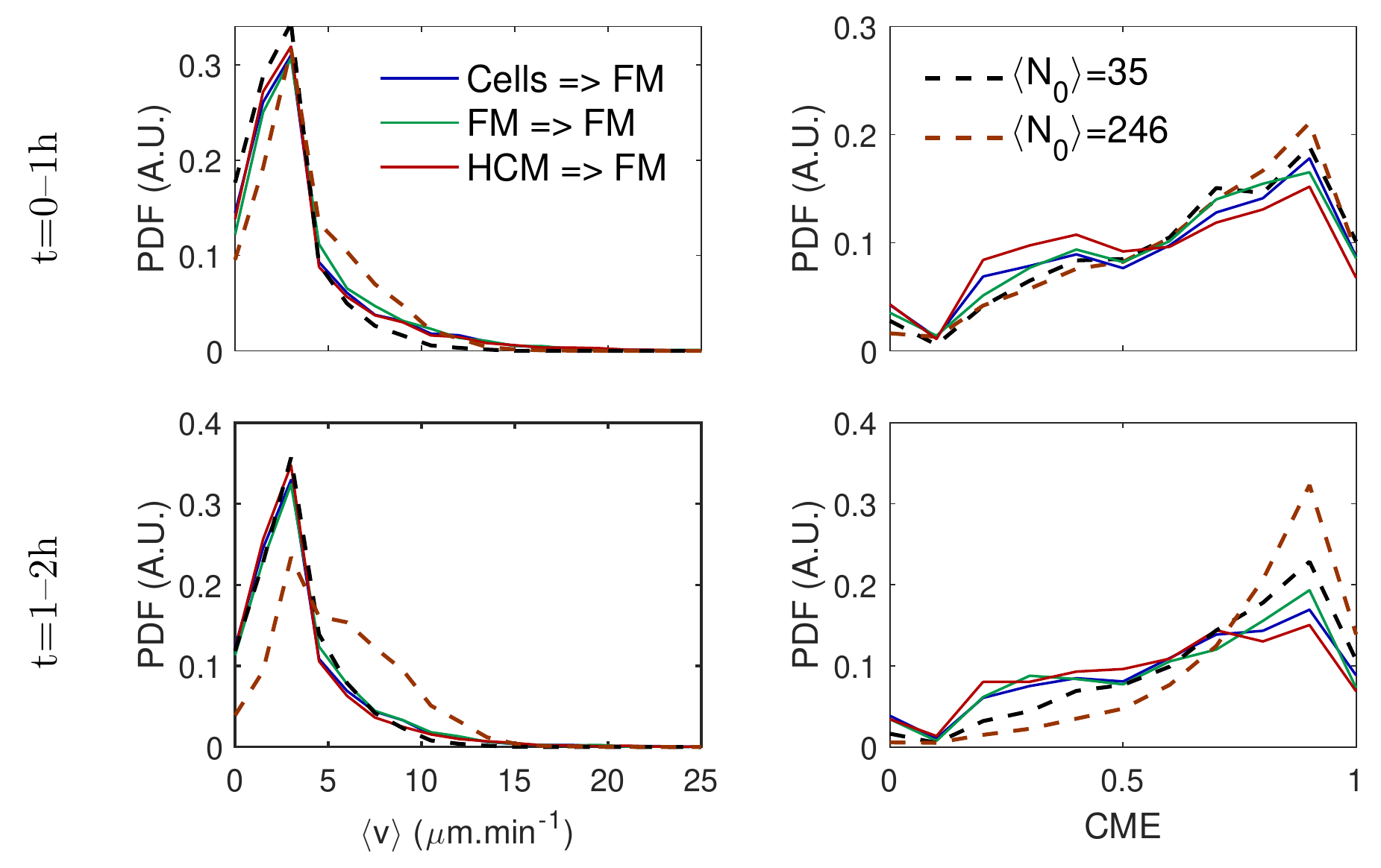}

\caption{Modifications of the surface by the cells do not affect cell motility.\\
  Distribution of speed (left) and CME with $\Delta t=5\,$min (right)
  computed from trajectories with $\delta t=1\,$min, gathered for all times
  $t=0-1$\,h (top) and $t=1-2$\,h (bottom) with various surface treatments (solid lines) or in spreading colonies (dashed lines). Cells $\Rightarrow$ FM: cells
  were seeded at high density, let adhere for 45\,min and washed out
  before adding fresh cells in fresh medium (FM) for imaging. FM
  $\Rightarrow$ FM: the sample dish was filled with FM for 45\,min
  before cells and FM were added for imaging. HCM $\Rightarrow$ FM:
  the sample dish was filled with HCM for 45\,min before cells and FM
  were added for imaging.}
\label{fig:sup_trails}
\end{figure*}

We also tested the hypothesis of cell-cell communication through
  the deposition of chemicals on the surface as follows. First, the
  sample dishes were treated by letting cells adhere on their entire surface at high density
  for 45\,min. Then the cell layer was washed out and fresh cells were
  added at a low, homogeneous density with fresh medium for imaging
  and the single cells were tracked after 45\,min adhesion. The controls
  include prior incubation with FM or HCM but no cell layer. If
  cell-cell communication through deposited trails on the surface
  affects the motility, there should be measurable differences between
  those conditions. Conversely, since the system is homogeneous and isotropic and the cells
  undergo almost no cell-cell contact, no effect of geometry or
  interaction is expected to interfere with those prior treatments.
  Yet, the probability density functions (PDFs) of speed and CME in the
  three conditions are well overlaid, showing no effect of a putative
  chemical deposition mechanism. Those PDFs compare very well with
  those measured in the low density colonies, while they differ
  from the high density colonies, especially
  around the peak of $\langle v_r\rangle(t)$ at $t=1$--2\,h 
  (Supp. Fig.~\ref{fig:sup_trails}).

Taken together, these results show that a large-scale communication
using secreted or deposited molecules is very unlikely to be at the
origin of the density-dependent spreading at short times. In
consequence, the interactions behind this effect must be local,
whether mediated by actual cell-cell contacts or by rapidly degraded,
locally accumulated chemicals. Although the latter is not
inconceivable, it is less likely at stake and can be described as an
``effective contact'' interaction.

\clearpage
\section{Increase in single cell persistence}

\begin{figure*}[h!]
\centering
\includegraphics[scale=0.6]{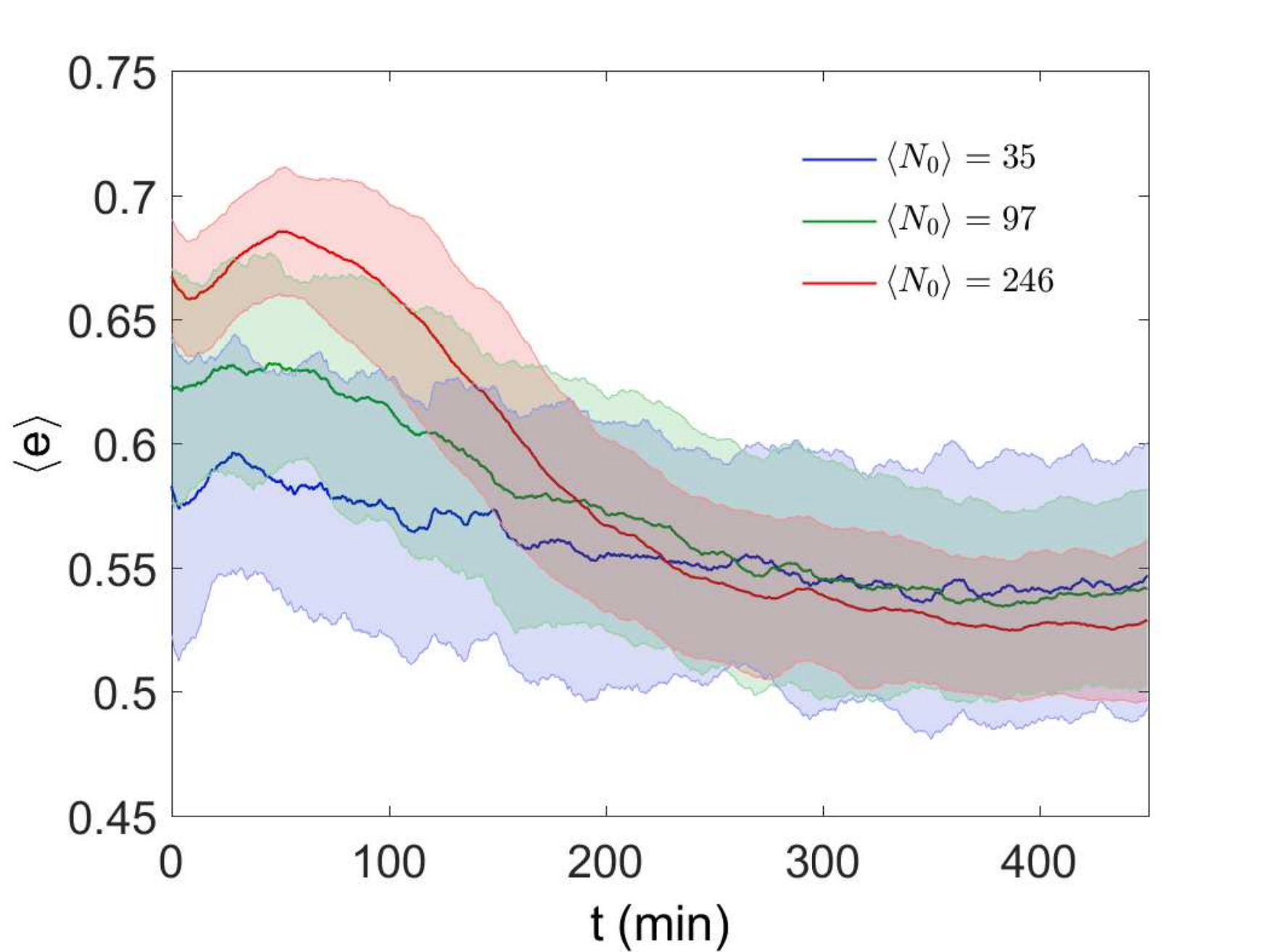}

\caption{Cells tend to be more elongated at high density.\\
  Time evolution of the average eccentricity of the cell shapes'
  obtained by fitting ellipses for different initial densities
  $\langle N_0\rangle$. The error bars are the standard deviations of the $e$ distributions.}
\label{fig:sup_ecc}
\end{figure*}

In \textit{Dictyostelium discoideum}, the more persistent cells are
also the more elongated ones, while the ones that exhibit rounded
shapes usually move in a more random
fashion~\cite{Gole2011,Heid2004}. A way to measure the elongation of
cells is to fit their shape with ellipses and compute the eccentricity
$e=\sqrt{1-\frac{b}{a}}$, where $a$ and $b$ are respectively the long
and short axes of the ellipse. The eccentricity varies between $0$ for
a circle and $1$ for an infinitely elongated ellipse. For instance,
$e=0.5$ corresponds to $a/b=1.33$, $e=0.7$ to $a/b=1.96$ and $e=0.9$
to $a/b=5.26$. We measured the colony average of $e$ as a function of
time for different cell densities (Supplementary Figure
\ref{fig:sup_ecc}). At early times, $\langle e\rangle$ goes up with
cell density. It then relaxes to $\langle e\rangle\approx 0.55$ for
all $\langle N_0\rangle$.  Although this elongation seems to be partly
defined prior to the release of the stencil (see $\langle e\rangle$ at
$t=0$), a peak still appears in the densest condition, concomitantly
with the peaks in $\langle v_r\rangle$ and $\phi_2$.

\begin{figure*}[h]
\centering
\includegraphics[scale=0.65]{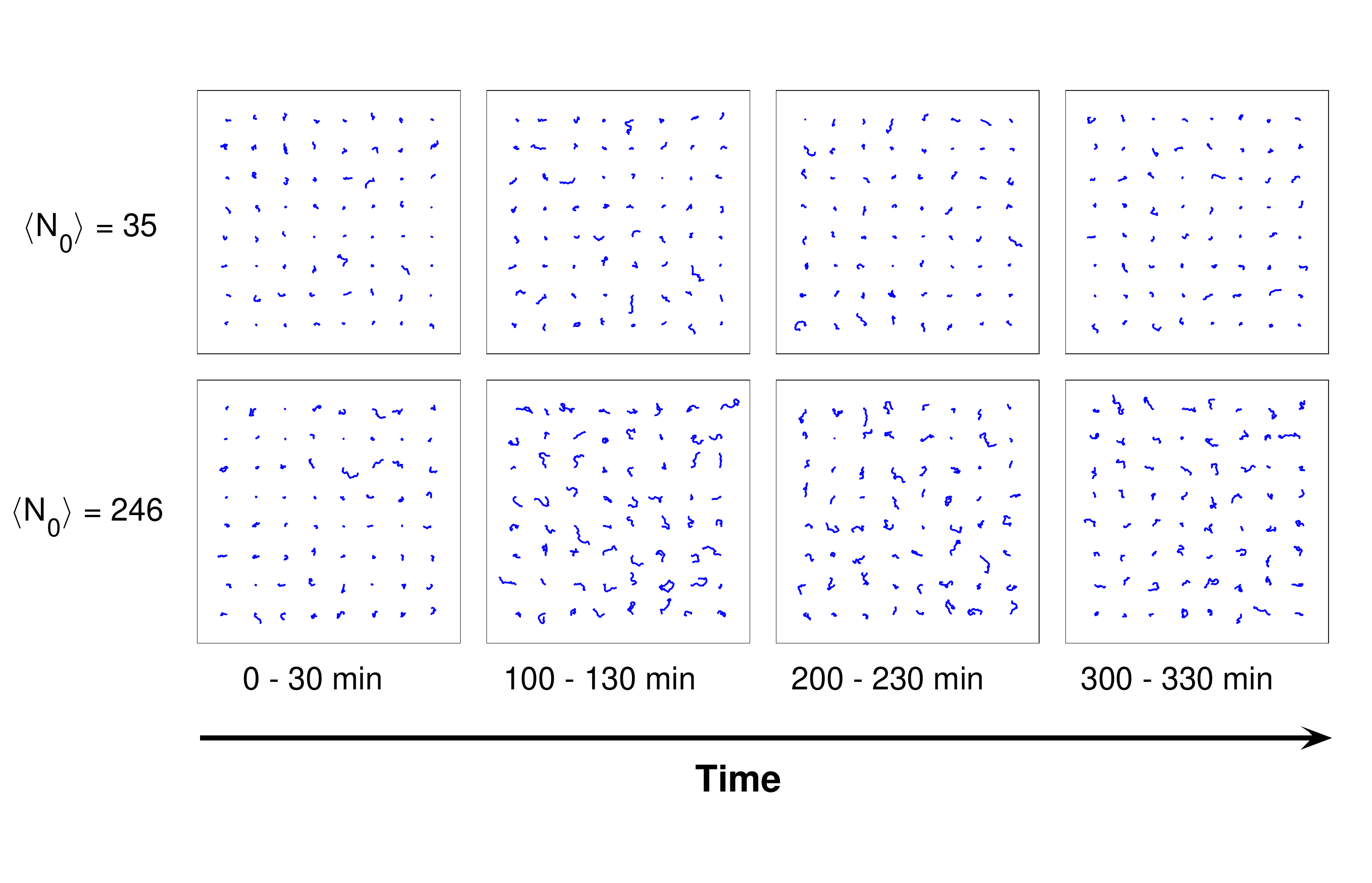}

\caption{Increase in single cell persistence.\\
  Example of single cell trajectories from the lowest
  ($\langle N_0\rangle = 35$) or highest ($\langle N_0\rangle = 246$)
  cell density experiments, at times $t=0-30\,\minute$,
  $t=100-130\,\minute$, $t=200-230\,\minute$,
  $t=300-330\,\minute$. The persistence is seen to increase
  transiently around $t=100\,\minute$ in the high density case, while
  at low density it does not change markedly in time. Grid spacing:
  $200\,\micro$m.}
\label{fig:sup_trajs}
\end{figure*}

As for the persistence of cell trajectories, it increases transiently
with no detectable pre-set dependence on density. To illustrate this
fact, we show randomly sampled trajectories -- all of the same
$30\,$min duration -- taken from experiments with low
($\langle N_0\rangle=35$) or high ($\langle N_0\rangle=246$) cell
densities at different times (Supplementary Figure
\ref{fig:sup_trajs}). At early times, most trajectories show a low
persistence for both densities. This changes in the high density case
where trajectories appear ``unfolded'' around $t=100\,$min,
demonstrating the existence of runs with persistence time comparable
to the $30\,$min path duration. At later times, the motion becomes
again less persistent.

\clearpage
\section{Bimodal persistent motion}

\rev{
We found two modes in the decay of the velocity direction auto-correlation
function (VDACF). As mentioned, either those two modes arise from two separate
populations of cells, each having a single correlation time, or every
single cell exhibit both modes. The first hypothesis is ruled out since
it does not allow changes in the weights of the modes. In the second
hypothesis, the motion of each cell follows a process with two characteristic
times: this is the case of bimodal motion\cite{Potdar2010,Maiuri2015,Metzner2015},
but other models have this same property\cite{Selmeczi2005}. Here we focus on
the first option because it offers the simplest interpretation of a change
in the weights $c$ and $c'$ in Equation (\ref{eq:corr_biexp}). In the model
proposed by Selmeczi \textit{et al.}\cite{Selmeczi2005}, it would mainly
involve tuning their $\alpha$ parameter, which is the strength of a memory in
a modified O.U. process: the underlying principle -- regulation of the relative
importance of two time-scales -- is the same, but the formulation involves
a higher degree of complexity which seems unnecessary here.}

\paragraph{Model.}
We first characterise analytically a simple model of bimodal
persistent motion.  Assume a particle moves in the plane with velocity
of constant magnitude $v_0$.  Its orientation is subject to rotational
diffusion, but with a coefficient that alternates between two values
$\Dru$ and $\Drd$.  The times spent in mode 1 and 2 are both
exponentially distributed, with mean $\tauu=\lamu^{-1}$ and
$\taud=\lamd^{-1}$ respectively\footnote{The value of $D_r$ is thus a
  Telegraph process.}.  What are the properties of such a random
motion?

The essential quantity  is the velocity direction autocorrelation function
\be
\C(t'-t) =  \langle  \vcu(t')\cdot \vcu(t)\rangle  =  \langle \cos \left[ \theta(t') - \theta(t)  \right]\rangle,  
\ee
where $\vcu = (\cos \theta, \sin \theta)$ is the direction of motion. 
Let's introduce the probability densities  $p_{i=1,2}(\theta,t)$  to be in mode~$i$ with orientation $\theta$ at time~$t$, 
they are governed by
\begin{linenomath}
\begin{subequations}
\label{eq:p1p2}
\begin{align}
\partial_t \pru &= \Dru\, \partial^2_{\theta\theta}\pru + \lamd \prd - \lamu \pru,  \qquad  \pru(\theta,0)=\phi_1 \delta(\theta),    \label{eq:p1p2a}\\ 
\partial_t \prd &= \Drd\,\partial^2_{\theta\theta} \prd - \lamd \prd + \lamu \pru,  \qquad  \prd(\theta,0)=\phi_2    \delta(\theta), \label{eq:p1p2b}
\end{align}
\end{subequations}
\end{linenomath}
where $\phi_1 = \lamd/(\lamu + \lamd)$ (resp.
$\phi_2 = \lamu/(\lamu + \lamd)$) is the fraction of time spent in
mode~1 (resp. mode~2), and we have taken for initial orientation
$\theta=0$.  Now, introducing
$p(\theta,t)=\pru(\theta,t)+\prd(\theta,t)$, $\C(t)$ can be expressed
as \be \C(t) = \int_{-\pi}^{\pi}\, d\theta \, p(\theta,t) \cos\theta.
\ee The system \eqref{eq:p1p2} can be solved using Laplace transforms
for time and Fourier series for orientation, yielding \be \C(s) =
\frac{ (\lamu + \lamd)^2 + \lamu (s+\Dru) + \lamd (s+\Drd)} {(\lamu +
  \lamd) \left[ (s+\Dru)(s+\Drd) +\lamu (s+\Drd) + \lamd (s+\Dru)
  \right]}, \nonumber \ee where variable $s$ is the Laplace variable.
Going back to time domain, this expression translates into the sum of
two exponentials \be \C(t) = c\, e^{-\gamma t} + c'\, e^{-\gamma' t},
\label{eq:corr_francois}   
\ee
with the notations
\begin{linenomath}
\begin{subequations}
\label{eq:corr_params}
\begin{align}
\kappa^2  &= (\Dru+\Drd + \lamu + \lamd)^2 - 4 (\Dru \Drd + \Dru \lamd + \Drd \lamu ),            \\
\kappa'^2 &= (\Dru-\Drd + \lamu - \lamd)^2 + 4 \lamu  \lamd,                                      \\
\gamma    &= (\kappa +\Dru+\Drd + \lamu + \lamd )/2,                                              \\
\gamma'   &= \gamma - \kappa,                                                                     \\
c=1-c'    &= - \frac{\left(\Dru-\Drd\right) \left(\lamu-\lamd\right)+\left(\lamu+\lamd\right) \left(\lamu+\lamd -\kappa' \right)}{2 \kappa  \left(\lamu+\lamd\right)}.                           
\label{eq:c_prime}
\end{align}
\end{subequations}
\end{linenomath}
It can be shown that the following inequalities hold \be c,c'
\geqslant 0, \qquad \Drd \leqslant \gamma' \leqslant \Dru < \gamma,
\ee showing that both terms are always decaying, and that the slowest
relaxation is intermediate between $\Dru$ and $\Drd$.

In the limit $\Dru \rightarrow \infty$, {\it i.e.} when all
directional persistence is lost in mode~1, all expressions greatly
simplify \be \gamma = \Dru, \quad \gamma'=\Drd+\lamd, \quad c'=
\frac{\lamu}{\lamu + \lamd}, \ee where only the first term in the
expansion has been retained.  In this case, $\C(t)$ exhibits a rapid
drop with characteristic time $\sim \Dru^{-1}$, followed by a slowest
decay whose constant $\gamma'=\Drd+\lamd$ is independent of $\Dru$ and
whose prefactor $c'$ is the fraction of time spent in mode 2.  On
further assuming that mode 2 involves ballistic motion ($\Drd=0$),
then $\gamma'=\lamd$, giving access to the mean duration of mode 2.
In this particular case, the relationship between the biexponential
form of correlation function and the model parameters is simple.  This
is not so, however, in the general case, and accordingly we have
resorted to a fitting procedure.

\begin{figure*}[ht!]
\centering
\includegraphics[scale=0.6]{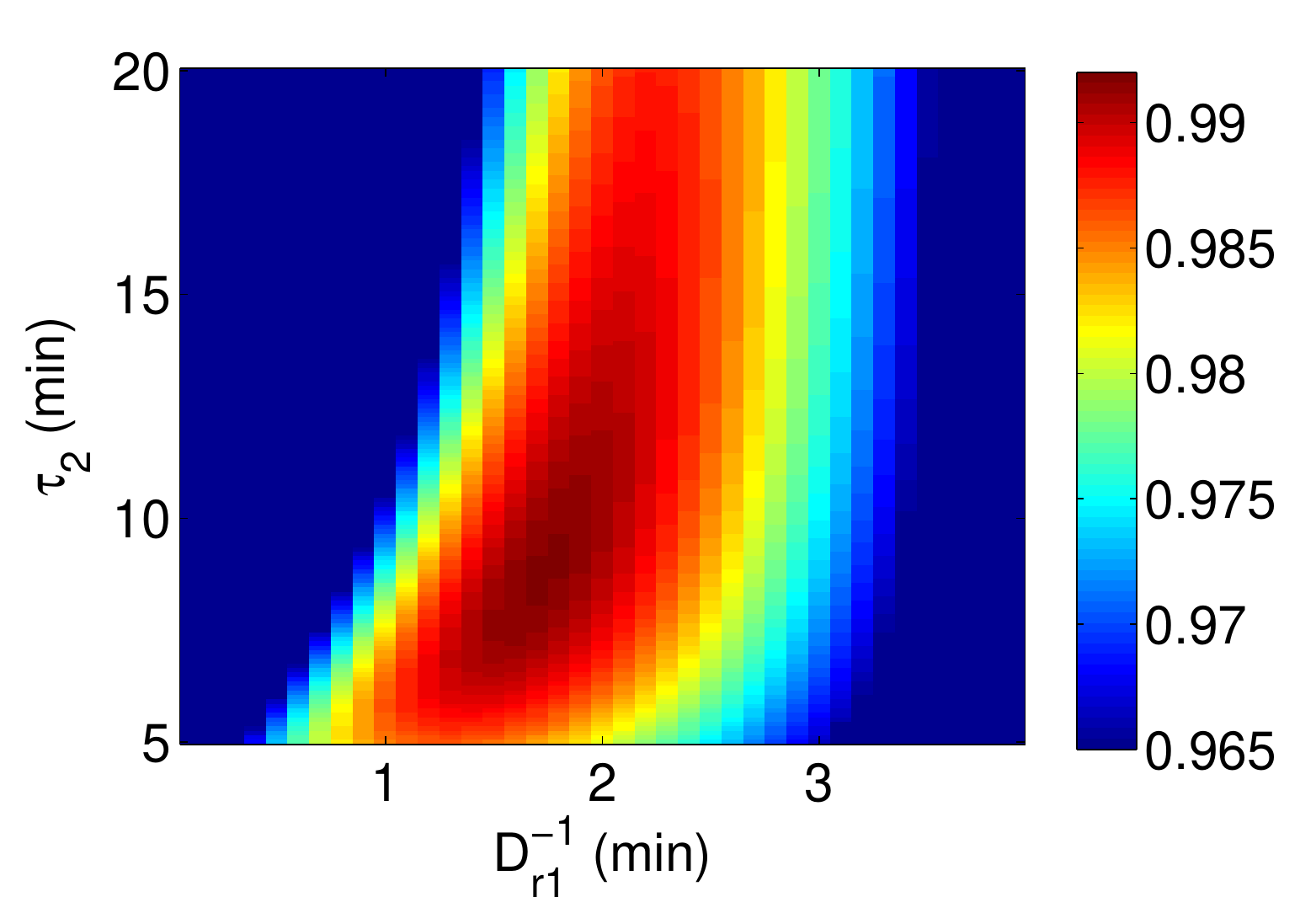}

\caption{Parameter space for the fitting procedure.\\
  The average $R^2$ computed from 25 experimental curves of
  correlation functions exhibits a clear peak at \rev{
    $\Dru^{-1}=1.7\,$min and $\taud=8.6\,$min.  } Here, the
  time-windows for $t\geqslant200\,$min were not considered in order
  not to overweight the long-term behaviours. Using the complete set
  of 39 curves slightly moves the peak of $R^2$ but then
  $\Dru^{-1}=2\,$min and $\taud=10\,$min remain excellent estimates.}
\label{fig:sup_R2}
\end{figure*}

\paragraph{Fitting procedure -- experimental data.}
We use the expressions above (\ref{eq:corr_francois} \&
\ref{eq:corr_params}) to fit all experimental correlation functions.
Since mode~2 is assumed ballistic, $\Drd=0$.  The parameters $\Dru$
and $\taud$, considered as intrinsic properties of the cells, are
common to all curves and are thus heavily constrained.  The free
parameters remaining for each curve are $\tauu$, and an additional
constant prefactor that allows $C(t=0)$ to differ from unity.
 In practice, we first varied systematically the values of $\Dru$
  and $\taud$.  For each couple, we fitted the complete set of 15
  $C(t)$ curves (5 first time-windows for each of the 3 density
  conditions) with the two mentioned free parameters. We used the
  value of $R^2$ averaged on these 15 curves to estimate the quality
  of the fit for a given couple ($\Dru$, $\taud$).  This yielded
  unambiguously the optimal values $\Dru^{-1}=1.7$\,min and
  $\taud=8.6$\,min (Supplementary Figure \ref{fig:sup_R2}). Then we
  measured $\tauu$ by fitting again all the curves using these
  optimised fixed parameters.  

In using the model, we have tacitly assumed that at each time, the
population of cells is ``equilibrated'' between the two modes.  This
is reasonable since the residence time in each mode ($\tau_1$ and
$\tau_2=10\,$min) remains smaller than the time scale over which
$\lamu$ (or the density) significantly varies (around 100\,min).

\begin{figure*}[ht!]

\centering
\includegraphics[scale=0.6]{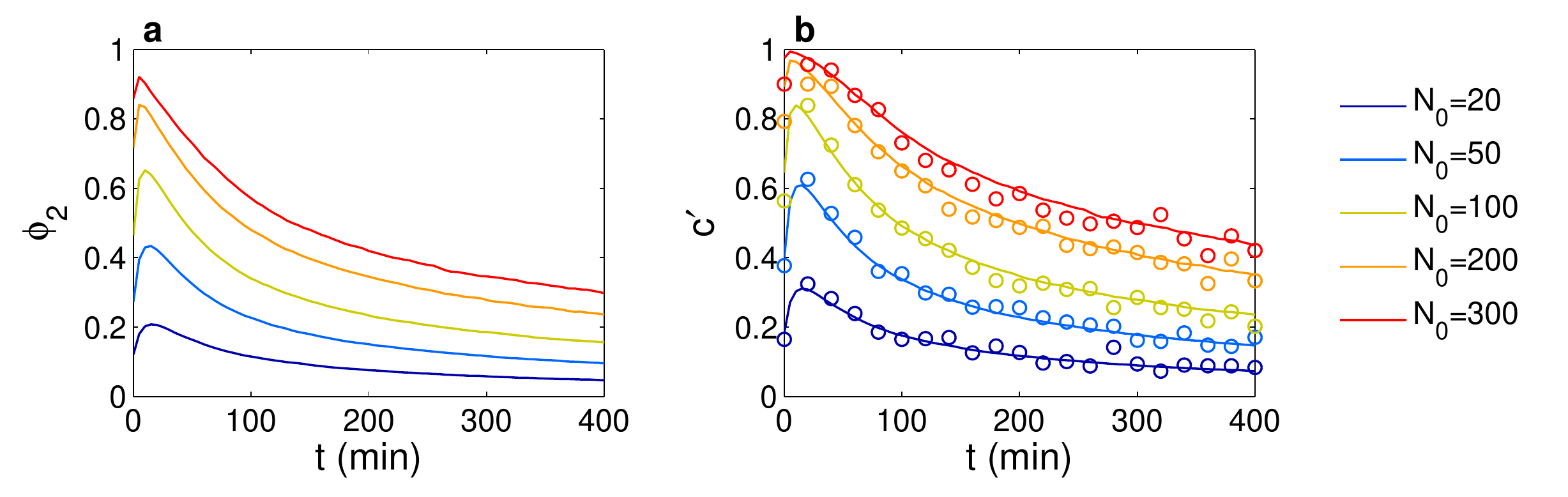}

\caption{Simulations -- Results of the fitting of the velocity autocorrelation functions applied to simulations with CEL.\\
  (a) Proportion of cells in mode 2.  (b) Prefactor $c'$ of the
  longest relaxation in the correlation function. Theoretical
  prediction from $\phi_2$ (---) and measurement from the correlation
  functions (o).}
\label{fig:sup_corr_sim}
\end{figure*}

\paragraph{Fitting - simulations with CEL.}
We computed the proportion of particles in persistent mode $\phi_2$ at
all times (direct output from the simulations, Supplementary Figure
\ref{fig:sup_corr_sim}a) and the velocity autocorrelation functions at
various times and for all particle numbers $N_0$. From the latter, we
could extract the parameters from a fit with expression
(\ref{eq:corr_francois}).  In particular we compared the obtained $c'$
values to the theoretical predictions (\ref{eq:c_prime})
(Supplementary Figure \ref{fig:sup_corr_sim}b). They are in close
agreement with each other, showing that the hypothesis of
equilibration between the two modes through $\lambda_1$, under the
control of collisions, is reasonable.

\section{Analysis of pair collisions}

\subsection{Detection of contacts.}

\begin{figure*}[ht!]
\centering
\includegraphics[scale=0.75]{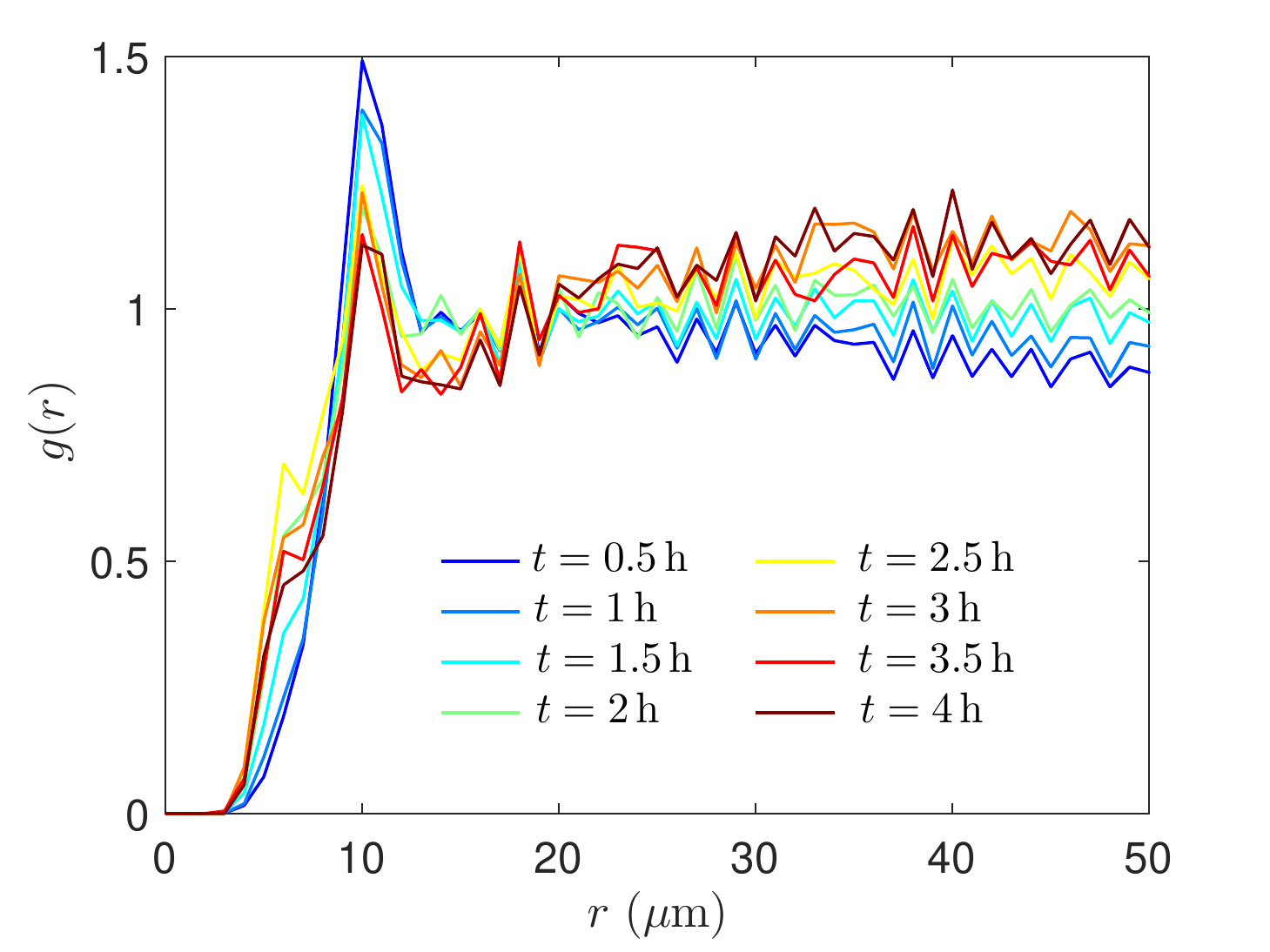}

\caption{Radial distribution function, averaged over all the
  experiments, at various time points.}
\label{fig:sup_rad_dist}
\end{figure*}

We study here the effect of cell-cell contacts on the trajectories. To
that end, we need to find a way to disentangle this effect from other
biases related to the occurence of contacts: For instance, contacts
are more likely to happen in dense areas, where the behaviour is known
to be different on average from sparser zones. For that reason, and to
avoid increasing the number of spurious underlying parameters, we only
focused on two-body interactions.

A cell-cell contact was defined by two cell positions being closer
than a distance $d_{max}$. If the cells remain close longer than one
time frame, the collision time $t_0$ is defined as the time at which
the cell-cell distance is minimal. Moreover, since our measurements
cannot be instantaneous -- \textit{ie} they are based on pieces of
trajectories extending over several time frames -- we selected only
those collisions involving two cells that did not encounter any other
collision for a time $t_{free}$ before \textit{and} after $t_0$, hence
reducing drastically the number of exploitable data.

Contrary to the simulations, for which the interaction radius is
well-defined, the choice of $d_{max}$ is not trivial in experiments
because cell shapes and sizes are distributed. Thus, to fix $d_{max}$
we first measured the radial distribution function $g(r)$
(Supp. Fig.~\ref{fig:sup_rad_dist}). The profiles, computed at
different times, all show a marked peak at $r=10\,\micro$m. As a
consequence, we chose a slightly larger $d_{max}=11\,\micro$m to
detect reliably the contacts between cells.

\subsection{Angular deflection}

\begin{figure*}[ht!]
\centering
\includegraphics[scale=0.5]{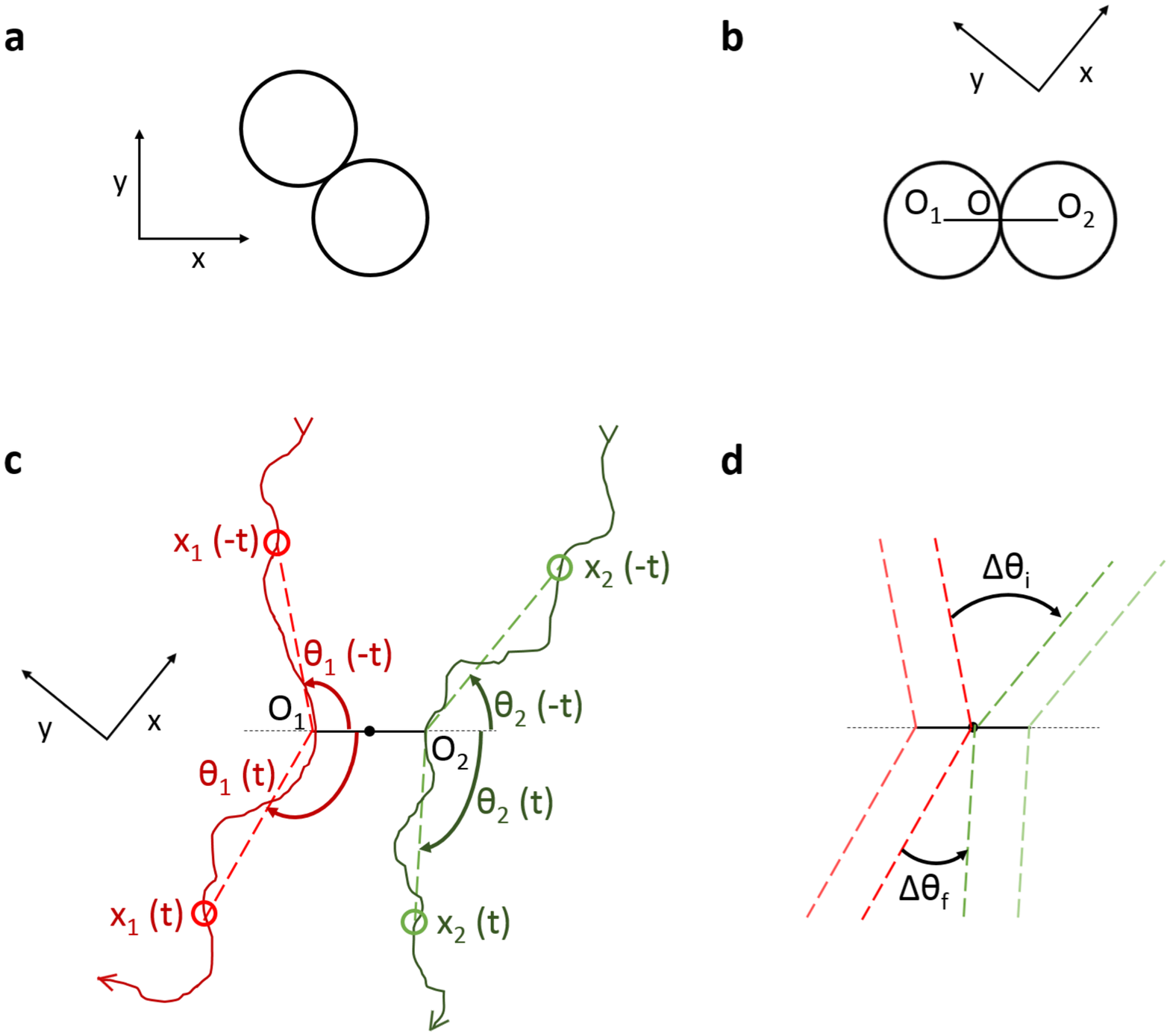}

\caption{Sketch of the angle definition.\\
  (a) A cell-cell contact is detected in the frame of the picture.
  (b) The frame is rotated so that the $(O_1OO_2)$ axis becomes the new
  $x$-axis, where $O_{i}$ is the position of cell $i$ and O is the
  barycentre of the collision.  (c) The incidence (resp. scattering)
  angle $\theta_{1/2}(-t)$ (resp. $\theta_{1/2}(t)$) is measured from
  the cell's position at $-t$ respective to its position at the time
  of the contact.  (d) The incidence (resp. scattering) angle separation
  $\Delta\theta_i$ (resp. $\Delta\theta_f$) is the difference between
  the two directions of motion of the cell couple.}
\label{fig:sup_collision}
\end{figure*}

\begin{figure*}[ht!]
\centering
\includegraphics[scale=0.5]{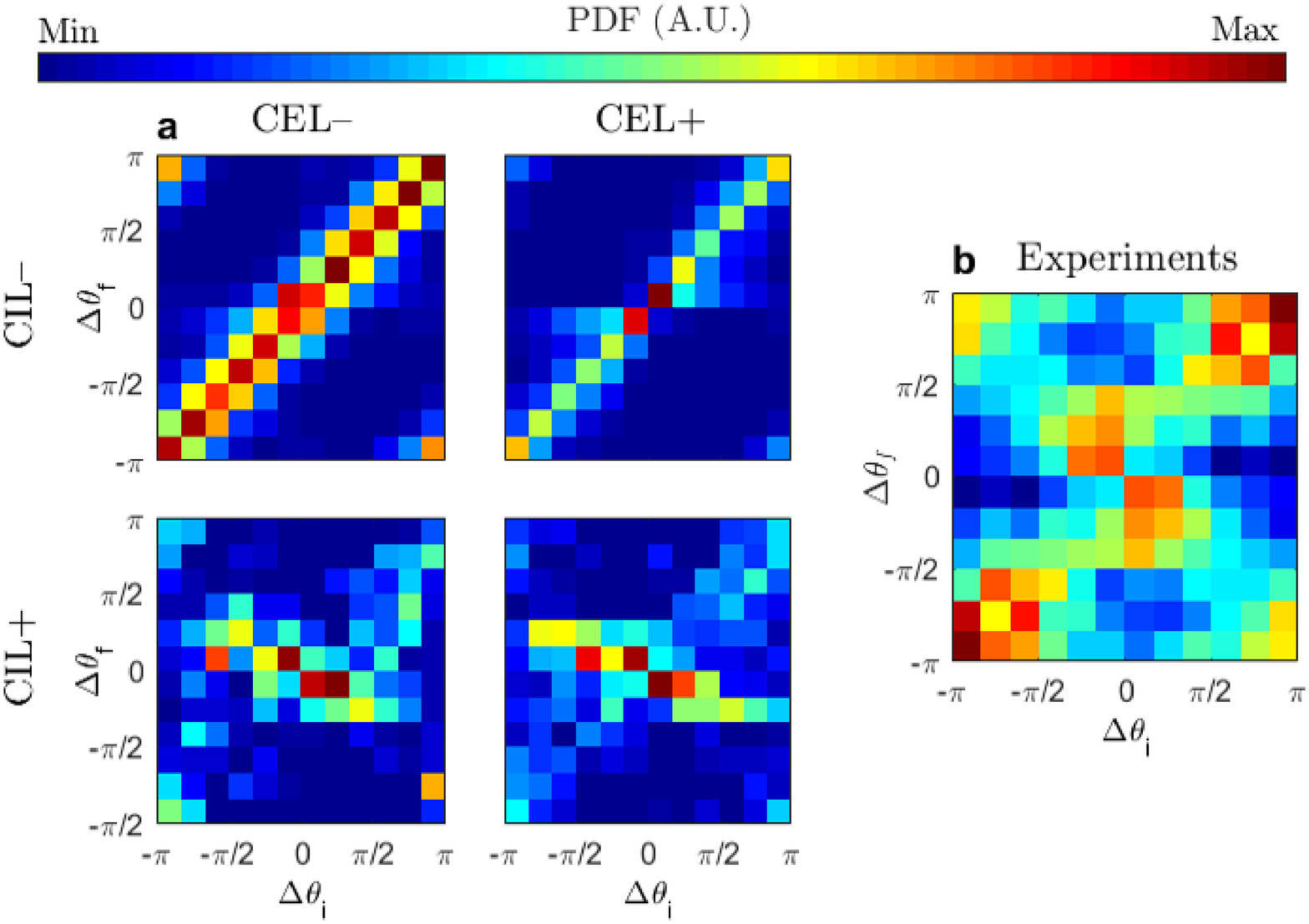}

\caption{Angular deflection at collision in simulations and experiments.\\
  $\mathbb{P}(\Delta\theta_f|\Delta\theta_i)$ for collisions detected with
  $d_{max}=11.2\,\micro$m, $t_{free}=3\,$min, at
  $t_{f/i}=\pm1\,$min. (a) Simulations without (CIL--) or with (CIL+) repulsive torque
  and monomodal ($D_r^{-1}=10\,$min, CEL--) or bimodal (CEL+) motion.
  (b) Experimental data gathered from all experiments.}
\label{fig:sup_angles}
\end{figure*}

\begin{figure*}[ht!]
\centering
\includegraphics[scale=0.9]{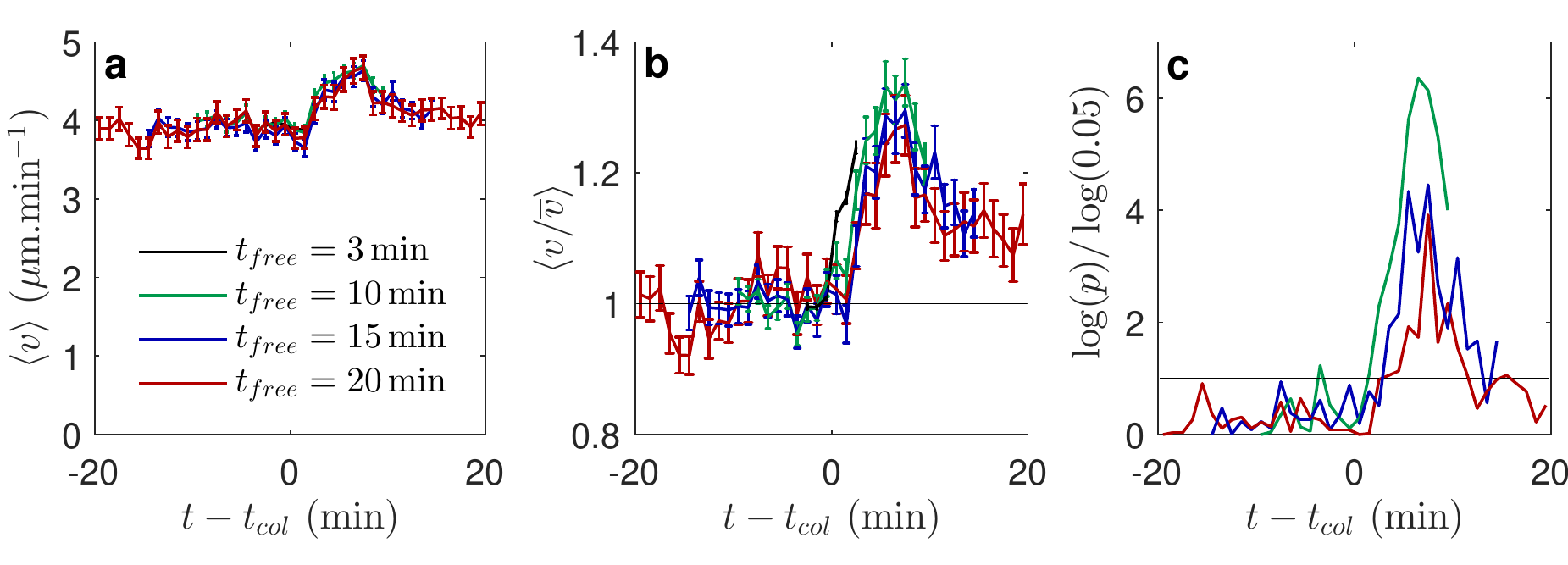}

\caption{Increase in cell speed after collisions.\\
  (a) Average speed before and after a single collision, using
  smoothed trajectories with $\delta t=1\,$min and $t_{free}=3$
  (black), 10 (green), 15 (blue) and 20\,min (red). Mean$\pm$SEM for
  $n=6578, 376, 232$ and $153$ cell pairs respectively.  (b) Average
  normalised speed $v/\overline{v}$, where $\overline{v}$ is the basal
  speed of a single cell before and after collisions. Same
  $\delta t$, $t_{free}$ and number of contacts as in (a).  (c)
  Logarithm of the $p$-value obtained from the Kolmogorov-Smirnov test
  against the null hypothesis that $v(t)/\overline{v}$ is distributed
  with the same PDF as $v(-t_{free})/\overline{v}$. Values above 1
  denote when the null hypothesis can be rejected with more than 95\%
  confidence.}
\label{fig:sup_coll_speed}
\end{figure*}

We studied in details the statistics of incidence and scattering
angles, both in simulations and experiments. To that end, we detected
the collisions with $t_{free}=3\,\minute$ before and after the
collision. As depicted in Fig.~\ref{fig:sup_collision} the cell
coordinates were rotated so that the $(O_1OO_2)$ axis is the new
$x$-axis, where $O_1$ and $O_2$ are the positions of cell 1 and 2 at
collision time, and $O$ is the center of $[O_1O_2]$. We then measured
the mean angles of motion $\theta_i(\pm t)$, where $i\in\{1,2\}$ and
$t\in\{1,2,3\}\,\minute$, defined as the angle between $(O_iX_i(t))$ and
$(OO_i)$ and computed $\Delta\theta_{i}=\theta_2(-t)-\theta_1(-t)$
and $\Delta\theta_{f}=\theta_2(t)-\theta_1(t)$, the incidence and scattering
angles respectively. Trying to decipher the effect of the collision,
we are especially interested in the relative scattering $\Delta\theta_f$
as a function of $\Delta\theta_i$. In Fig.~\ref{fig:sup_angles}, we show
the heat maps for the probability densities $\mathbb{P}(\Delta\theta_f|\Delta\theta_i)$.

In simulations without CIL, the probability distribution
$\mathbb{P}(\Delta\theta_f|\Delta\theta_i)$ is concentrated around the
main diagonal $\Delta\theta_f=\Delta\theta_i$. It means that the
relative direction of motion remains unaffected by the collision,
hence that the particles cross ``without seeing'' each other (in terms
of direction of motion).  It is consistent with the simulation rule
that the interaction only involves a pushing force during the contact,
but the intrinsic direction of motion is not modified.

The results are completely different in simulations with CIL
(introduced in the form of a repulsive torque). Most of the
probability concentrates in a zone where $\dth_i$ and $\dth_f$ have
opposite signs, a signature of the angular repulsion.

In the experimental data, two distinct behaviors seem to be present:
There is both a concentration of probability around both diagonals,
$\dth_f=\dth_i$ denoting crossing, and $\dth_f=-\dth_i$, a signature
of specular reflection. In particular, $\dth_f=-\dth_i$
seems more likely for small $\dth_i$, when the incident trajectories
are close to being parallel, while $\dth_f=\dth_i$ is
predominant near $\|\dth_i\|=\pi$.  These results support the idea
that even in the event of a collision, \textit{Dictyostelium
  discoideum} cells reorient smoothly so as to bypass their encounter:
at low incidence angle crossing is difficult while almost specular
reflection demands only a slight turn, and at higher incidence it is easier
to circumvent each other. It is different from the usual view of CIL,
according to which colliding cells reorient specifically away along the
contact axis. It could be related to the increased
probability for \textit{D. discoideum} cells to form new pseudopods in
the protruded area rather than along the cell body\cite{Bosgraaf2009},
acting together with a CIL-like inhibition of protrusions in the
cell-cell contact zone. In any case, they show that the cell-cell contacts
have no aligning effect on the direction of motion, and thus could not
generate a coherent motion that would be responsible for the increase of the colony spreading rate

\subsection{Contact enhancement of locomotion.}

\paragraph{Cell speed.}
Similarly as for the direction of motion, we measured the
instantaneous speed of cells undergoing a single collision in a frame
with free motion $t_{free}$ before and after the collision.  We found
that on average, the cells exhibit a transient increase of their speed
after collisions. Because of positional noise, this effect is not
completely clear using the experimental time frame $\delta t=20\,$s.
Yet it is better seen after smoothing the trajectories over
$\delta t=1\,$min, and varying $t_{free}$ from 3\,min to 20\,min:
$t_{free}=20\,$min provides a view of the long-term dynamics; with
shorter $t_{free}$, the complete time-frame is not accessible, but the
trend for $\langle v\rangle(t)$ is confirmed with much more statistics
(up to 6578 cell-cell contacts measured with $t_{free}=3\,$min).

The cells accelerate for $7-8\,$min following collisions, reaching an
average speed approximately 20\% higher than the ``basal'' average
speed, and then the speed decreases back to its basal value at a
similar rate.  When the speed of each single cell is normalised by its
own basal speed $\overline{v}$ (\textit{i.e.} its mean speed prior to
the collision), one even gets a 30\% increase on average. Although
these changes are relatively small compared to the standard deviation
of the speed distribution in the cell population, a statistical
analysis using the Kolmogorov-Smirnov test shows that the distribution
differs significantly from the distribution at $t=-t_{free}$ only for
$\sim12\,$min after collisions.

\begin{figure*}[ht!]
\centering
\includegraphics[scale=0.9]{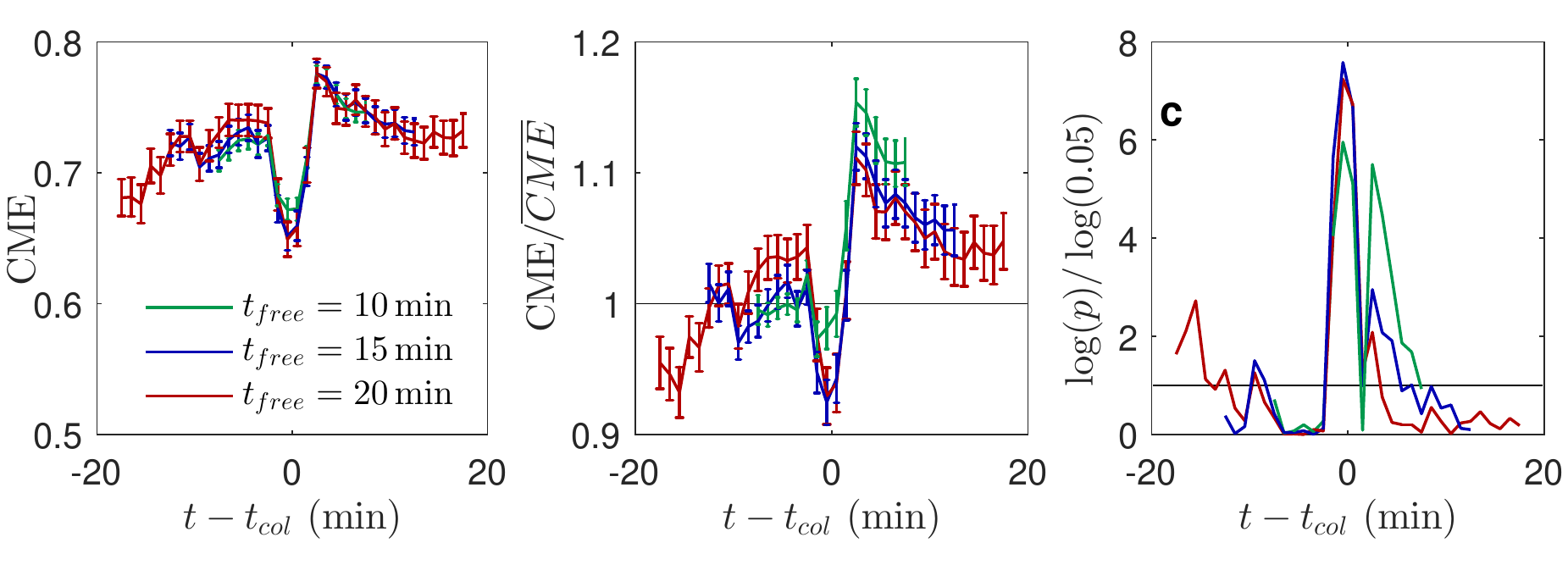}

\caption{Increase in CME after collisions.\\
  (a) Average CME before and after a single collision, using smoothed
  trajectories with $\delta t=1\,$min and $t_{free}=10$ (green), 15
  (blue) and 20\,min (red). Mean$\pm$SEM for $n=376, 232$ and $153$
  cell pairs respectively.  (b) Average normalised CME
  CME$/\overline{\text{CME}}$, where $\overline{v}$ is the basal speed
  of a single cell, before and after collisions. Same $\delta t$,
  $t_{free}$ and number of contacts as in (a).  (c) Logarithm of the
  $p$-value obtained from the Kolmogorov-Smirnov test against the null
  hypothesis that CME$(t)$ is distributed with the same PDF as
  CME$(-t_{free})$. Values above 1 denote when the null hypothesis can
  be rejected with more than 95\% confidence.}
\label{fig:sup_coll_CME}
\end{figure*}

\paragraph{Persistence time.}
Using again smoothed trajectories with $\delta t=1\,$min to make the
patterns of evolution more apparent, we repeat the analysis on the CME
computed for $\Delta t=5\,$min (see Methods). This revealed a CME drop
around the contacts, as well as a subsequent transient increase
(Supp. Fig.~\ref{fig:sup_coll_CME}a-b). Both effects seemingly arise
from significant changes in the distribution of the CME
(Supp. Fig.~\ref{fig:sup_coll_CME}c). The depression is probably
related to changes in direction during the contact. The increase,
between 10 and 15\% for single cells on average, shows that the
persistence of the motion is enhanced for at least a few minutes after
a cell-cell contact.

\begin{figure*}[ht!]
\centering
\includegraphics[scale=0.6]{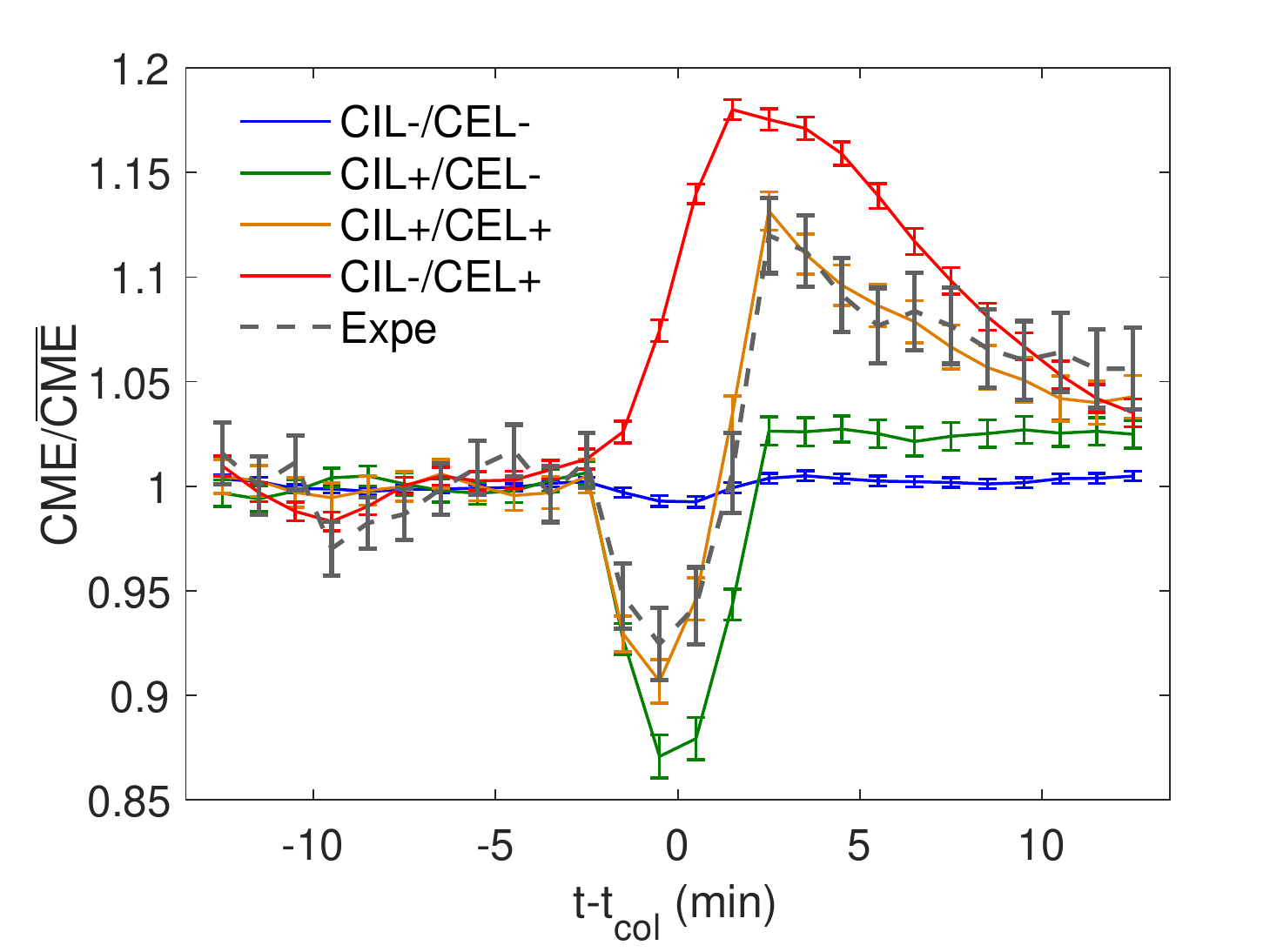}

\caption{Single-cell-normalised CME: comparison of various simulation conditions with experimental data. Mean$\pm$SEM, $t_free=15\,$min.}
\label{fig:sup_coll_CME_simus_xp}
\end{figure*}

\paragraph{Sensitivity of CME measurements to the presence of CEL.}
While in the case of the speed the measurement is direct, the CME is
only a proxy to estimate whether the trajectories are more or less
persistent on a given time-scale $\Delta t$. As a consequence, we also
tested our CME-based analysis of collisions on the simulation data. It
makes apparent that a drop in CME around 10\% is the signature of CIL
(in the sense of turning upon contact), while a transient increase in
CME of 10--20\% is observed only in the presence of CEL
(Supp. Fig.~\ref{fig:sup_coll_CME_simus_xp}).

Finally, the experimental curve aligns closely with that of the
CIL+/CEL+ simulations
(Supp. Fig.~\ref{fig:sup_coll_CME_simus_xp}). Without proving that the
cells follow the precise rules implemented in the simulations, it
shows that CIL and CEL seem to be necessary to account for the
experimental collision data.

\section*{Supplementary movies}

\paragraph{Supplementary movie 1}
Long-time spreading of a colony with high cell density.

\paragraph{Supplementary movie 2}
Short-time ($200\,\minute$) dynamics of a colony with low cell density. $N_0=18$.

\paragraph{Supplementary movie 3}
Short-time ($200\,\minute$) dynamics of a colony with high cell density. $N_0\simeq275$.

\paragraph{Supplementary movie 4}
Simulation of active particles at low density undergoing CEL. $N_0=20$.

\paragraph{Supplementary movie 5}
Simulation of active particles at high density undergoing CEL. $N_0=200$.

\end{document}